\documentclass[a4paper,11pt]{article}
\pdfoutput=1 % if your are submitting a pdflatex (i.e. if you have
\usepackage{jheppub} % for details on the use of the package, please
\usepackage[T1]{fontenc} % if needed
\usepackage{natbib}
\usepackage{here}
\usepackage{float}
\usepackage{multirow}
\usepackage{graphicx}
\usepackage{subcaption}
\usepackage{multirow}
\usepackage{bbm}
\usepackage{cleveref}

\usepackage{tikz}
\usetikzlibrary{decorations.pathmorphing}
\usetikzlibrary{decorations.markings}
\usetikzlibrary{positioning, shapes, arrows}
\usetikzlibrary{calc}
\usepackage[compat=1.0.0]{tikz-feynman}

\tikzset{
quarkar/.style={postaction={decorate}, decoration={markings,mark=at position .5 with {\arrow[#1]{latex}}},very thick},
quark/.style={postaction={decorate}, decoration={markings}, very thick},
scalar/.style={dashed,postaction={decorate}, decoration={markings,mark=at position .5 with {\arrow[#1]{latex}}}},
gluon/.style={decorate,decoration={coil,amplitude=3pt, segment length=4.7pt, pre length=.01cm, post length=.01cm}},
gluont/.style={decorate,decoration={coil,amplitude=3pt, segment length=3.50pt, pre length=.01cm, post length=.01cm}},
photon/.style={decorate, decoration={snake, segment length=6pt, amplitude=1.8pt,  pre length=1pt, post length=1pt}},
photontop/.style={/tikz/preaction={draw=white,line width=5pt}, decorate, decoration={snake, segment length=6pt, amplitude=1.8pt,  pre length=1pt, post length=1pt}},
cut/.style={red, very thick, dashed},
}

\usepackage[utf8]{inputenc}

\usepackage{hyperref}
\usepackage{bbold}

\usepackage{graphicx,epsf}
\usepackage{amsmath,amsfonts,amssymb,amsbsy}

\usepackage{hyperref}
\usepackage{xcolor}
\hypersetup{
    colorlinks=true,
    allcolors={blue!60!black}
}

\usepackage{multirow}
\usepackage{here}
\usepackage{stackrel}
\usepackage{tikz}
\usetikzlibrary{calc}
\usetikzlibrary{positioning} 

\usepackage{ifdraft}
\usepackage[obeyFinal]{easy-todo}

\usepackage{placeins}
\usepackage{slashed} % For Dirac slashed notation

\usepackage{booktabs}
\usepackage{adjustbox}

\newcommand{\draftnote}[1]{\todo{#1}}

\newcommand{\be}{\begin{equation}}
\newcommand{\ee}{\end{equation}}

\newcommand{\G}{\Gamma}

\newcommand{\bea}{\begin{eqnarray}}
\newcommand{\eea}{\end{eqnarray}}

\newcommand{\cI}{\mathcal{I}}

\newcommand{\bm}{\bar{m}}

\newcommand{\cla}{{\rm cl}}
\newcommand{\hdelta}[0]{\hat{\delta}}

\def\G#1{${\mathcal O}(G^#1)$}

\def\spa#1.#2{\left\langle#1\,#2\right\rangle}
\def\spb#1.#2{\left[#1\,#2\right]}
\def\spash#1.#2{\spa{\smash{#1}}.{\smash{#2}}}
\def\spbsh#1.#2{\spb{\smash{#1}}.{\smash{#2}}}
\def\sand#1.#2.#3{%
\left\langle\smash{#1}{\vphantom1}^{-}\right|{#2}%
\left|\smash{#3}{\vphantom1}^{-}\right\rangle}
\def\sandpp#1.#2.#3{%
\left\langle\smash{#1}{\vphantom1}^{+}\right|{#2}%
\left|\smash{#3}{\vphantom1}^{+}\right\rangle}
\def\sandpm#1.#2.#3{%
\left\langle\smash{#1}{\vphantom1}^{+}\right|{#2}%
\left|\smash{#3}{\vphantom1}^{-}\right\rangle}
\def\sandmp#1.#2.#3{%
\left\langle\smash{#1}{\vphantom1}^{-}\right|{#2}%
\left|\smash{#3}{\vphantom1}^{+}\right\rangle}

\newcommand*{\Caravel}{\textsc{Caravel}}
\newcommand*{\pySecDec}{py{\textsc{SecDec}}}

\newcommand*{\LiteRed}{\texttt{LiteRed}}

\newcommand*{\PolyLogTools}{\texttt{PolyLogTools}}

\title{
Gravitational Bremsstrahlung in Black-Hole Scattering at $\mathcal{O}(G^3)$: 
Linear-in-Spin Effects
}

\preprint{PSI-PR-23-47}

\author[1]{Lara Bohnenblust}
\emailAdd{lara.bohnenblust@uzh.ch}
\author[1,2]{Harald Ita}
\emailAdd{harald.ita@psi.ch}
\author[3]{Manfred Kraus}
\emailAdd{mkraus@fisica.unam.mx}
\author[1]{Johannes Schlenk}
\emailAdd{johannes.schlenk@psi.ch}

\affiliation[1]{ ICS, University of Zurich, Winterthurerstrasse 190, 8057 Zurich, Switzerland}
\affiliation[2]{ Paul Scherrer Institut, CH-5232 Villigen PSI, Switzerland}
\affiliation[3]{Departamento de F\'{i}sica Te\'{o}rica, Instituto de F\'{i}sica, \\ Universidad Nacional Aut\'{o}noma de M\'{e}xico, Cd. de M\'{e}xico C.P. 04510, M\'{e}xico}

\abstract{
We compute the far-field time-domain waveform of the gravitational waves
produced in the scattering of two spinning massive objects. The results
include linear-in-spin ($S$) couplings and first-order gravitational
corrections ($G^3$), and are valid for encounters in the weak-field regime.
Employing a field-theory framework based on the scattering of massive scalar
and vector particles coupled to Einstein-Hilbert gravity, we derive results for
leading and the next-to-leading spectral waveforms. 
We provide analytic expressions for the required scattering data, which
include trees, one-loop amplitudes and their cuts.
The expressions are extracted from numerical amplitude evaluations with the
\Caravel{} program, using analytic reconstruction techniques applied in the classical limit.
We confirm a recent prediction for infrared physics of the classical
observable, and observe the surprising appearance of a ultraviolet singularity, which
drops out in the far-field waveform.
}

\begin{document}
\maketitle
\newpage

%--------------------------------------------------------------------------------
%--------------------------------------------------------------------------------
\section{Introduction}
%--------------------------------------------------------------------------------
In recent years, the groundbreaking detection of gravitational
waves~\cite{LIGOScientific:2016sjg,LIGOScientific:2016dsl} has opened a new
frontier in astrophysics, providing new and revolutionary means to explore the
universe and unlock profound insights into the nature of spacetime itself.
Future upgrades of the existing gravitational-wave observatories of the
LIGO-Virgo-KAGRA collaboration~\cite{KAGRA:2021vkt} and planned observatories such
as Cosmic Explorer~\cite{Reitze:2019iox}, Einstein
Telescope~\cite{Punturo_2010} and LISA~\cite{LISA:2017pwj}, operating in lower
frequency ranges will explore new types of systems including fly-bys,
captures, eccentric configurations and high spin. 
To fully exploit the physics potential of the observatories, it will be
important to increase the theoretical precision and scope of gravitational
waveform predictions in the near future. 

The central challenge in obtaining a waveform of binary systems is the
non-linearity of the gravitational interaction coupled with the dependence on
multiple physical scales.  
Present theoretical predictions rely on numerical
relativity~\cite{KAGRA:2021vkt} and the effective-one-body (EOB) formalism
\cite{Buonanno:1998gg,Buonanno:2000ef}, based on input from post-Newtonian (PN)
dynamics \cite{Blanchet:2013haa,Schafer:2018kuf}, the gravitational self-force
formalism \cite{Mino:1996nk,Quinn:1996am}, the non-relativistic
general-relativity (NRGR)
\cite{Goldberger:2004jt,Goldberger:2006bd,Goldberger:2009qd,Porto:2012as,Blanchet:1994ez} or 
the weak-field Post-Minkowskian (PM) approximation 
\cite{Einstein:1940mt,Bertotti:1956pxu, Kerr:1959zlt, Bertotti:1960wuq, Portilla:1979xx,
Westpfahl:1979gu, Portilla:1980uz, Bel:1981be, Westpfahl:1985tsl,
Damour:2016gwp}.

In contrast, we focus here on a first-principle waveform computation in the PM expansion, which we
derive to next-to-leading order in the Newton constant $G$, i.e. to order \G3.
This perturbative approach allows us to systematically compute dynamics and observables
of binary black-hole systems and is applicable in a weak-field regime and at large eccentricities.  
The PM approach has rapidly advanced in recent years starting from its initial
application in General Relativity
\cite{Bertotti:1956pxu,Kerr:1959zlt,Bertotti:1960wuq,Portilla:1979xx,Westpfahl:1979gu,Portilla:1980uz,Bel:1981be,Westpfahl:1985tsl,Ledvinka:2008tk}.
In particular the recent use of modern field-theory methods has led to
impressive high-order \G3~\cite{Bern:2019nnu} and \G4~\cite{Bern:2021dqo,Bern:2021yeh,Dlapa:2021npj,Kalin:2022hph,Dlapa:2022lmu}
predictions for classical quantities, and has already helped to improve
resummation of the effective-one-body (EOB)
Hamiltonians~\cite{Khalil:2022ylj,Rettegno:2023ghr}. It has further fueled the developement of the heavy-mass effective field theory (HEFT)~\cite{Damgaard:2019lfh,Aoude:2020onz,Haddad:2020tvs,Brandhuber:2021kpo},
 worldline EFTs~\cite{Kalin:2020fhe,Liu:2021zxr,Dlapa:2021npj,Kalin:2022hph,Dlapa:2022lmu,Kalin:2020mvi} and eikonal approaches~\cite{DiVecchia:2021bdo,DiVecchia:2022nna,DiVecchia:2022piu}.
As far as the waveform is concerned, methods are available to
directly link this observable to scattering amplitudes (KMOC)
\cite{Kosower:2018adc,Maybee:2019jus,Cristofoli:2021vyo} and to worldline
quantum field theory (WQFT)
~\cite{Jakobsen:2021smu,Jakobsen:2021lvp,Mougiakakos:2021ckm}. By now, early classical results 
~\cite{Kovacs:1977uw,Kovacs:1978eu}
have been reproduced for the scattering of two Schwarzschild black holes at
leading-order \G2 , and were recently extended to the next-to-leading order
\G3~\cite{Herderschee:2023fxh,Brandhuber:2023hhy,Georgoudis:2023lgf,Elkhidir:2023dco}.
The importance of the so-called cut contribution to this observable for imposing
classical causality was pointed out in \cite{Caron-Huot:2023vxl}. This contribution will
likely be crucial for a comparison with the multipolar PM
waveform~\cite{Bini:2023fiz}.  Recently such a comparison, including cut
contributions, was reported for the limit of soft
radiation~\cite{Georgoudis:2023eke}.

A further motivation for this work is to consider spin effects in the waveform
observable, as required for describing astrophysical black holes.
The foundational work on the PM treatment of spin
\cite{Bini:2017xzy,Vines:2017hyw,Bini:2018ywr,Bern:2020buy} has sparked many 
conceptional and computational developments, and has exposed new
links between field theory and GR \cite{Bautista:2022wjf,Bautista:2023sdf}. 
Focusing on perturbative computations, spin corrections to two-body dynamic
have by now been obtained up to \G4 in worldline approaches
\cite{Jakobsen:2023ndj,Jakobsen:2023hig,Liu:2021zxr} and at \G3 in field theory
\cite{FebresCordero:2022jts}.
The latter obtains spin corrections from massive field theories
\cite{Bern:2022kto,Aoude:2022thd,Aoude:2023vdk,Chiodaroli:2021eug,Cangemi:2022abk,Cangemi:2022bew,Ochirov:2022nqz,Cangemi:2023ysz,Alessio:2023kgf}
via effective field theory \cite{Bern:2020buy} or via scattering amplitudes
using a generalization of the KMOC
formalism~\cite{Maybee:2019jus,Menezes:2022tcs}.
However, how to construct the right field theory, that mirrors the properties of
Kerr black holes, remains an open question. 
At low spin orders minimally coupled theories correctly capture the spin
couplings of macroscopic objects \cite{Vaidya:2014kza}. This `spin
universality' property, was validated for the minimally coupled vector theory
up to quadratic-in-spin multipoles at \G3 \cite{FebresCordero:2022jts}.
Turning to the waveform observable, \G2 quadratic spin effects were obtained
in ref.~\cite{Jakobsen:2021lvp} and to higher-spin
order~\cite{DeAngelis:2023lvf,Brandhuber:2023hhl,Aoude:2023dui}, recently.
Beyond the leading PM order, the spin-dependent memory effect is known at
\G3~\cite{Aoude:2023dui}.

The goal of this article is to present the spectral waveform emitted by two
colliding black holes including \G3 corrections, and to include linear-in-spin
effects.  We provide analytic results in terms of momentum-transfer variables
and in frequency space. Furthermore, we present exemplary plots of the
gravitational waveform in the time and position-space domain, and demonstrate
the impact of the cut contribution in the gravitational waveform. 
The waveform observable is computed from first principles in QFT following the
KMOC formalism \cite{Kosower:2018adc,Cristofoli:2021vyo}.  We exploit the
classical limiting procedure and the detailed analytic and physical
understanding of
refs.~\cite{Herderschee:2023fxh,Brandhuber:2023hhy,Georgoudis:2023lgf,Elkhidir:2023dco},
but let QFT do its work.  We provide an independent computation of the
necessary Feynman integrals and one-loop scattering data.  In addition, we
provide the cut contribution, that allows us to obtain the full non-spinning
waveform at \G3.

Building on these results, we extend this study by including linear-in-spin
effects, which we extract from a one-loop computation including 
massive vectors and scalars. 
This approach exploits the relation between minimally coupled vector-field theory and spinning
point-particles
\cite{Howe:1988ft,Bastianelli:2005vk,Bastianelli:2005uy,Vaidya:2014kza}.
Recently, a similar approach has been pursued in the analysis of
the \G3 conservative dynamics quadratic in spin \cite{FebresCordero:2022jts}, 
which we build on.

We present a purely numerical approach to obtain analytic results to handle 
the multi-scale computation.
The KMOC formalism relates the waveform observable at \G3 to
tree-level and one-loop scattering amplitudes and their cuts.  
We obtain all scattering data with the \Caravel-program 
\cite{Abreu:2020xvt}, which implements exact numerical evaluations using modular arithmetic. 
The program allows to efficiently handle
the intricate gravitational coupling structure of the massive scalar and vector
fields.  At tree-level, the program yields numerical scattering
amplitudes through Berends-Giele recursions \cite{Berends:1987me}, 
using interaction vertices obtained with
\textsc{xAct}~\cite{Brizuela:2008ra,Nutma:2013zea}.
At loop-level, the program implements the numerical unitarity method
\cite{Ossola:2006us,Ellis:2007br,Berger:2008sj,Giele:2008ve,Ellis:2008ir} in
the variant of refs.~\cite{Ita:2015tya,Abreu:2017xsl,Abreu:2017hqn}, which
provides numerical values for the rational integral coefficients of a one-loop
integral basis.
We employ functional reconstruction techniques
\cite{Peraro:2016wsq,vonManteuffel:2014ixa} to obtain analytic expressions in
the classical expansion parameter and isolate the classical terms. The full
analytic expressions are obtained using the functional-reconstruction method
\cite{Peraro:2016wsq,Abreu:2017xsl} in finite-field arithmetic
\cite{vonManteuffel:2014ixa}.
An important aspect of computing spin corrections is the relation between 
polarization states of vector fields and classical spin variables.
We obtain this map, by introducing a form-factor decomposition~\cite{FebresCordero:2022jts}, which 
we then link to classical spin multi poles~\cite{Cangemi:2022bew}.
We anticipate that this setup will allow to obtain further higher-spin
corrections to the \G3 waveform, which we leave for future work.

We make a number of interesting observations. For instance, we confirm the recent predictions 
for the IR-singularities of the waveform
observable \cite{Caron-Huot:2023vxl} and link their result to a modification of the Weinberg's IR-theorem
\cite{Weinberg:1965nx}.
Furthermore, we observe an additional contribution to the $1/\epsilon$-pole in
the dimensional regulator in the cut contribution, which we attribute to the UV. 
The UV pole is shown to integrate to zero in the
Fourier transformation to position-space in the far-field asymptotic waveform.
In the light of the recent comparisons with previous results for the
gravitational waveform in the Post-Newtonian
expansion~\cite{Bini:2023fiz,Georgoudis:2023eke}, we provide a new input for the
scalar and for the linear-in-spin waveform at \G3.

The article is organized as follows.  In \cref{sec:notation}, we establish our
notation and present our conventions for spin operators and polarization states
of gravitons and massive vector fields.  Next, in \cref{sec:overview}, we give a
brief overview how the gravitational waveform can be related to scattering
amplitudes. In \cref{sec:amplitude}, we explain in detail how the
necessary scattering amplitudes are computed. We put special emphasis on the
description of our exact numerical approach to obtain amplitudes in the
classical limit. In \cref{sec:results}, we collect our analytical results and
discuss various analytical features we observe.  Furthermore, in
\cref{sec:fourier}, we elaborate the computation of the position-space waveform
and show some numerical results for the gravitational waveform.  Finally, in
\cref{sec:conclusions}, we give our conclusions and an outlook.

%--------------------------------------------------------------------------------
\section{Notation and conventions}
\label{sec:notation}
%--------------------------------------------------------------------------------

We study gravitational radiation in classical two-body scattering far from
the source. Some of the classical bodies are assumed to carry spin.
We employ a QFT approach to
describe the classical system by coupling matter fields  to Einstein-Hilbert
gravity,
\begin{align}\label{eqn:EFT}
{\cal L}= {\cal L}_{\rm EH} + {\cal L}_{\rm matter}\;.
\end{align}
The classical information is then extracted 
from quantum scattering amplitudes by considering them in a scaling limit,
referred to as the \textit{classical limit}.
For the Einstein-Hilbert Lagrangian we follow the conventions,
\begin{align}\label{eqn:gravityLagrangian}
{\cal L}_{\rm EH} = -\frac{2}{\kappa^2} \sqrt{|g|} R\;, 
\end{align}
where $g={\rm det}(g_{\mu\nu})$ and $R$ the is Ricci-scalar defined by
$R=R^{\rho}_{~\mu\rho\nu}\,g^{\mu\nu}$. Further details on the definition of the
Riemann-tensor are given in appendix
\ref{sec:FieldTheoryApp}.
We work in the `t Hooft-Veltman (HV) scheme of dimensional regularization using
$D=4-2\epsilon$ and setting the dimension of the tensor algebra $D_s$ to
the same value, i.e. $D_s=4-2\epsilon$. Furthermore, we work with weak
gravitational fluctuations around a flat background $\eta_{\mu\nu}$.  The
fluctuations are described by the graviton field $h_{\mu\nu}$ in the
decomposition,
\begin{align}
g_{\mu\nu} = \eta_{\mu\nu}+\kappa\, h_{\mu\nu}\,.
\end{align}
We work in the mostly-minus metric convention $\eta={\rm diag}\{1,-1,-1,-1\}$
and use the gravitational coupling $\kappa=\sqrt{32 \pi G}$ in
terms of the Newton constant $G=\hbar G_N/c^3$, where $c$ is the speed of light, $\hbar$ 
the Planck constant and $x^0=ct$. 

Massive point-like sources are described by minimally-coupled field theories of
a massive scalar $\phi$ or a vector-field $V_\mu$ with Lagrangians,
\begin{align}\label{eqn:matterLagrangian}
{\cal L}_{(\phi,m)}&=\frac{1}{2} \sqrt{|g|} \Big[ g^{\mu\nu}(\partial_\mu \phi) (\partial_\nu \phi) - m^2 \phi^2 \Big] \;,\\  
{\cal L}_{(V,m)}&= -\frac{1}{4} \sqrt{|g|} \Big[ g^{\mu \rho}g^{\nu \sigma}F_{\mu\nu} F_{\rho\sigma}  
        - 2 m^2 g^{\mu \nu}V_{\mu}V_{\nu}\Big] \,, \quad F_{\mu\nu}=\partial_\mu V_\nu - \partial_\nu V_\mu \;.
\end{align}
The massive objects are represented as fundamental states of field theories,
and we associate the objects' intrinsic angular momentum to the fields' spin
representations. The mass parameter $m$ in the Lagrangian corresponds to the
physical mass.

In the following, we study two types of systems: Scalar and spinning compact objects. The
scattering of spin-less point-particles is described by the matter Lagrangian,
\begin{align}\label{eqn:matterLagrangianScalarI}
{\cal L}_{\rm matter}^{(i)}={\cal L}_{(\phi_1,m_1)} + {\cal L}_{(\phi_2,m_2)} \;.
\end{align}
To describe spinning point particles, we take advantage of the
observation~\cite{Vaidya:2014kza} that the scattering amplitudes involving
minimally coupled spin-$s$ massive particles can be used to extract classical-spin 
multipoles up to $S^{2s}$, i.e. $2s$ powers of the classical spin $S$.
Therefore, we consider a massive vector field
and a massive scalar, given by 
\begin{align}\label{eqn:matterLagrangianScalarII}
{\cal L}_{\rm matter}^{(ii)}= {\cal L}_{(V,m_1)} + {\cal L}_{(\phi_2,m_2)}  \;.
\end{align}
Our conventions for creation and annihilation operators as well as
polarization states are collected in appendix \ref{sec:FieldTheoryApp}.

%%%%%%%%%%%%%%%%%%%%%%%%%%%%%%%%%%%%%%%%%%%%%%%%%%%%%%%%%%%%%%%%%%%%%%%%%%%%%%%%%%
\begin{figure}[t]
\centering
\includegraphics{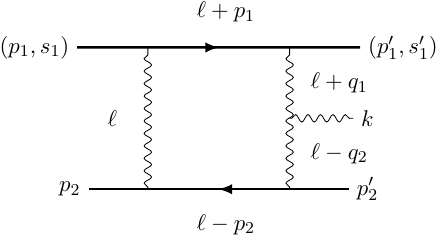}
\caption{Scattering process of two massive compact objects (solids lines),
emitting gravitational radiation (wavy line). Spin is indicated, as well as a
diagrammatic representation of the exchange of gravitons to order $\kappa^5$ in
perturbation theory.}
\label{fig:scattering}
\end{figure}
%%%%%%%%%%%%%%%%%%%%%%%%%%%%%%%%%%%%%%%%%%%%%%%%%%%%%%%%%%%%%%%%%%%%%%%%%%%%%%%%%%

In the above theories, we study the scattering process of four massive scalar
particles and one graviton, 
\begin{align}\label{eqn:procScalar}
{\rm i}: \quad & \phi_1(p_1) + \phi_2(p_2) \rightarrow \phi_1(p_1') + \phi_2(p_2') + h(k^{s})\,,
\end{align}
and the scattering of two massive vector particles and two scalar
particles,
\begin{align}\label{eqn:procVector}
{\rm ii}: \quad & V(p_1^{s_1}\, ) +\phi_2(p_2) \rightarrow V({p_1'}^{s_1'}\,) +\phi_2(p_2') + h(k^{s} )\,,
\end{align}
respectively.
Here the superscripts $s_1$ and $s_1'$ denote the spin quantum numbers of
the vector particles. The helicity of the graviton field $h$ is labeled
by the superscript $s$. The momenta $p_1$ and $p_2$ are understood
as incoming, and the final state momenta $p_1'$, $p_2'$ and $k$ are given
in the out-going convention. These kinematic conventions are summarized in fig.~\ref{fig:scattering}.

%--------------------------------------------------------------------------------
\subsection{Classical Scaling Limit}
\label{sec:classicalLimit}
%--------------------------------------------------------------------------------

We are concerned with the perturbative expansion in $G$ of the
classical scattering of two macroscopic rotating objects (large spin and mass)
and the emission of a classical, long wavelength gravitational wave. The
objects are assumed structureless, such as black holes.
We start our computation in a second-quantized field theory and will take
limits to retrieve such point-particle interaction in the weak-field expansion.

The kinematic properties of the corresponding scattering process are exposed in
the barred or soft
variables~\cite{Landshoff:1969yyn,Parra-Martinez:2020dzs,Luna:2017dtq},
\begin{equation}
\begin{split}
 p_1 &= \bm_1 u_1 + \frac{q_1}{2} \;, \qquad p_1^\prime = \bm_1 u_1 - \frac{q_1}{2}\;, \\
 p_2 &= \bm_2 u_2 + \frac{q_2}{2} \;, \qquad p_2^\prime = \bm_2 u_2 - \frac{q_2}{2}\;, \qquad k = q_1+q_2\;,
 \label{eqn:momenta}
\end{split}
\end{equation}
where $p_1^2 = p_1^{\prime 2} = m_1^2$, $p_2^2 = p_2^{\prime 2} = m_2^2$ and
$k^2 = 0$.
Here $u_i$ denote normalized $(u_i^2 =1)$ four velocities 
and the on-shell conditions of initial and final states imply $q_i
\cdot u_i$ = 0.  Similarly $k^2=(q_1+q_2)^2=0$ yields $q_1\cdot q_2 =
-(q_1^2+q_2^2)/2$\,.  The auxiliary mass parameters $\bm_i$ are related to the
physical masses through $\bm_i^2=m_i^2-q_i^2/4$.
We will also use the impact parameter four vectors $b_i$ which are Fourier
conjugate to the momenta $q_i$.
The interpretation of the momenta is as follows: The four-velocities $u_i$
represent classical velocities of the scattering objects, and $q_i$ 
(small) momentum transfer during their interaction. Part of the exchanged
momentum $k=q_1+q_2$ is emitted as radiation.

We use  seven Lorentz invariant inner products to
parameterize the kinematic space,
\begin{equation}\label{eqn:variables}
\begin{split}
  &y = u_1\cdot u_2\;,\qquad  \omega_1 = -u_1\cdot k \;,\qquad \omega_2=-u_2\cdot k \;,\\ 
  &q_1^2\;,\quad q_2^2\;, \qquad  \bm_1^2 \quad \mbox{and} \quad  \bm_2^2 \;.
\end{split}
\end{equation}
In the physical region of the phase space the invariants fulfill the constraints
\begin{equation}
 y > 1\;, \qquad q_i^2 < 0\;, \qquad \omega_i < 0\;, \qquad \bm_i > 0\;.
 \label{eqn:physregion}
\end{equation}

In the scattering process, the weak-field classical dynamic appears in the
following regime in field theory~\cite{Cheung:2018wkq,Bern:2019nnu}.
The weak-field expansion is a perturbative expansion in
\begin{align}
\frac{\kappa^2 m_i}{|\vec b_j|} \ll 1\,.
\end{align}
To ensure gravitational interaction of the masses we impose their effective
gravitational couplings to be large,
\begin{align}
    \kappa\, m_i \gg 1\,.
\end{align}
This is equivalent to the non-perturbative condition that the Schwarzschild radius  
of the masses is much larger than their Compton wave lengths, $\kappa^2\, m_i \gg 1/m_i$.
In terms of momentum variables the above constraints imply that the gravitational
interaction is long wavelength,
\begin{align}
\frac{k^\mu}{m_i} \sim \frac{q_j^\mu}{m_i} \ll 1  \,,
\end{align}
and weak,
\begin{align}
\kappa\, q_i^\mu \sim \kappa\, k^\mu \ll 1  \,,
\end{align}
compared to the matter-gravity interaction. Accordingly, in field-theory diagrams interactions 
with comparable momentum $\ell\sim q_i,k$ are
suppressed relative to the ones including matter lines.

Finally, the objects spin ($S_i$) is assumed to be macroscopic and
comparable in size to the angular momentum $\vec L$ of the system,
\begin{align}
|\vec S_i| \sim |\vec L_i| = |\vec p_i \times \vec b_i|\,. 
\end{align}
At the same time a perturbative expansion in spin requires the
ring-radius $a_i^\mu=S_i^\mu/m_i$ to be much smaller than the objects' Schwarzschild radius,
\begin{align}
 | \vec a_i | = |\vec S_i/m_i | \ll \kappa^2\, m_i   \,.
\end{align}
To implement the classical point-particle limit, we introduce the mass ratio
$q$,
\begin{align}
q=\frac{\bm_2}{\bm_1}\;,\quad \bm_1=\bm\;,\quad \bm_2=\bm\, q \;.       
\end{align}
The classical limit is then a scaling limit in
the variable $\bm\rightarrow \infty$ and amounts to performing a series
expansion around infinite mass $\bm$,
\begin{align}\label{eqn:classLimit}
\bm\rightarrow \infty \quad \mbox{with} \quad  y\,,q^2_i\,, q\,,\omega_i\,, \frac{S_i}{\bm} \,, \sqrt{\bm}\kappa = \mbox{fixed} \;.
\end{align}
An equivalent scaling limit via $\hbar$ counting \cite{Maybee:2019jus}, is obtained by rescaling 
all dimensionful quantities (including $\kappa$) simultaneous to the limit (\ref{eqn:classLimit}),
such that the mass-parameters $m_i$ remain fixed. For completeness we give this scaling transformation,
\begin{align}
\hbar \rightarrow 0 \quad \mbox{with} \quad \bm, y\,,\frac{q^2_i}{\hbar^2}\,, q\,,\frac{\omega_i}{\hbar}\,, \hbar S_i \,, \sqrt{\hbar} \kappa = \mbox{fixed} \;.
\end{align}
In our presentation we will rely on the former formulation
(\ref{eqn:classLimit}) of the limiting procedure.

In the classical limit, $L$-loop five-point scattering amplitudes including
four massive particles and one graviton have a well defined scaling in
$\bm$ and $\kappa$. For the order $n_s$ spin multipole contribution we have
\begin{align}\label{eqn:scalingM}
M^{n_s} \sim   \kappa^{3+2L}\, \bm^{4+L}\, (S/\bm)^{n_s}  \,.
\end{align}
The dependence on $\kappa$ follows from
Feynman rules. The dependence on $\bm$ and $S$ produces the
same leading scaling as the $M^{n_s=0}$ amplitude in the large-$\bm$ and $S/\bm={\rm
fixed}$ limit. 
The scalar amplitude scales like a fan-diagram \cite{Bern:2019nnu,Bern:2019crd,Cheung:2020gbf}, 
including $\bm^2$ for each matter-graviton vertex, and $1/\bm$
for massive propagators. We expect the scaling behavior,
\begin{align}\label{eqn:scalingMs}
M^{{\rm tree},n_s} \sim \kappa^3\, \bm^4\, (S/\bm)^{n_s} \,,\qquad 
M^{1-{\rm loop},n_s} \sim \kappa^5\, \bm^5\, (S/\bm)^{n_s} \,.
\end{align}
We refer to terms with the expected scaling dependence $\bm^{n_c}$ as
classical scaling. Slower growth $\bm^{n<n_c}$ will be called quantum and
faster growth $\bm^{n>n_c}$ hyper-classical. Hyper-classical scaling often
appears in intermediate computational steps and drops out in properly defined
observables. 

%--------------------------------------------------------------------------------
\subsection{Spin Operators}
\label{sec:spinOperators}
%--------------------------------------------------------------------------------
We compute scattering amplitudes which serve as building blocks of asymptotic
field-theory observables.  The classical spin dependence of the observables is
determined by making their dependence on spin operators manifest. In the
following, we briefly review the properties of spin operators.
 
The relativistic spin operator for a state with momentum $p$ and mass $m$
($p^2=m^2$) is the Pauli-Lubanski operator (see e.g. \cite{Sexl:1976pg}), which
is given as
\begin{align}\label{eqn:PLOperator}
\mathbb{S}_{\mu}=\frac{1}{2m}\epsilon_{\mu\nu\alpha\beta}\, p^\nu \mathbb{M}^{\alpha\beta}  \,,
\end{align}
for the Lorentz-group generators $\mathbb{M}^{\rho\sigma}$,
\begin{align}
\big[\mathbb{M}^{\mu\nu} , \mathbb{M}^{\rho\sigma} \big] &=  i
\big( \eta^{\nu\rho}\,\mathbb{M}^{\mu\sigma} 
- \eta^{\mu\rho}\, \mathbb{M}^{\nu\sigma} 
- \eta^{\nu\sigma}\, \mathbb{M}^{\mu\rho} 
+ \eta^{\mu\rho}\, \mathbb{M}^{\mu\sigma}\big)\,,
\end{align}
and where $\epsilon_{\mu\nu\alpha\beta}$ is the Levi-Civita tensor with
$\epsilon^{0123}=1$.

As we are working with spin-$1$ vector fields we only need the Lorentz-group
generators in the corresponding vector representation. We will specialize to
this case from now on and the generators explicitly read
\begin{align}\label{eqn:vectorRep}
(\mathbb{M}^{\mu\nu})^\alpha\,_{\beta} &= 
i \big(\eta^{\mu\alpha} \delta^\nu_{~\beta} -   \eta^{\nu\alpha} \delta^\mu_{~\beta} \big) \ \,.
\end{align}
The Pauli-Lubanski operators are transverse,
\begin{align}
\mathbb{S}_{\mu}\, p^\mu =0\;,
\end{align}
due to the anti-symmetry of $\epsilon_{\mu\nu\rho\sigma}$. Furthermore, they
are also transverse, viewed as tensors in their representation indices,
\begin{align}\label{eqn:transversality}
    (\mathbb{S}_{\mu})^\alpha_{~\beta} \, p^\beta=
    p_\alpha (\mathbb{S}_{\mu})^\alpha_{~\beta} = 0\,.
\end{align}
This property follows from the vector representation of the Lorentz generators
(\ref{eqn:vectorRep}) and their contraction with
$p^\nu\epsilon_{\mu\nu\rho\sigma}$.

For completeness we collect the commutator algebra of the Pauli-Lubanski
operators
\begin{align}
\big[\mathbb{S}^{\mu}, \mathbb{S}^{\nu}\big] &= -i\epsilon^{\mu\nu\rho\sigma} 
\mathbb{S}_{\rho} 
\frac{p^\sigma}{m}
\,.
\end{align}
In particular, for a state at rest $p=(m,0,0,0)$, the $\mathbb{S}^i$ form the
$so(3)$ Lie algebra of the respective little group. For this momentum choice,
$\mathbb{S}_0$ vanishes and the momentum $p$ is in the kernel of
$\mathbb{S}^i$.  The polarization states for a fixed momentum form irreducible
representations of the Pauli-Lubanski operators. 

Finally, we will require the projection operator
\begin{align}\label{eqn:projector}
\mathbb{P}^\alpha_{\,\beta}=\delta^\alpha_{~\beta} - \frac{p^\alpha p_\beta}{m^2}\,,
\end{align} 
which projects into the subspace transverse to the timelike momentum $p$. 
The Pauli-Lubanski operators $\mathbb{S}^\mu$, the projector 
$\mathbb{P}^\alpha_{~\beta}$, as well as the massive polarization states
$\varepsilon_{v\mu}(p)$ (with $\varepsilon_{v}(p)\cdot p=0$) all are
non-vanishing in the subspace transverse to $p$.

To make manifest the observables' dependence on spin, we will use a form-factor
decomposition which expresses amplitudes in terms of a complete basis of spin
operators and the transverse projector. In fact, the vector representation
operators
\begin{align}\label{eqn:tensorBasis}
\mathbb{P}^\alpha_{~\beta}\,,\quad  (\mathbb{S}^\mu)^\alpha_{~\beta}\, \quad\mbox{and}\quad   
(\mathbb{S}^{\{\mu}  \mathbb{S}^{\nu\}})^\alpha_{~\beta} ,
\end{align}
form an over-complete basis of rank two, transverse tensors, which can be
verified in the rest frame $p=(m,0,0,0)$. Here we take the indices $\mu$
and $\nu$ as labels and the representation indices $(\alpha,\beta)$ as the
mentioned tensor indices.
Furthermore, we defined the symmetric-traceless tensor
\begin{equation}
 \mathbb{S}^{\{\lambda}\mathbb{S}^{\kappa\}} \equiv 
 \frac{1}{2}\left(\mathbb{S}^\lambda \mathbb{S}^\kappa + \mathbb{S}^\kappa \mathbb{S}^\lambda \right) - 
 \frac{1}{3} (\mathbb{S}\cdot \mathbb{S})\, \mathbb{P}^{\lambda\kappa}\;.
\end{equation}
%
%--------------------------------------------------------------------------------
\subsection{Vector-boson States}
\label{sec:vectorStates}
%--------------------------------------------------------------------------------

An important fact is that the spin operators are non-vanishing in
the transverse space. Consequently, we set up the form-factor decomposition with 
respect to polarization states which take values in the same transverse space, 
that is polarization states associated to the same momentum.

To make this structure manifest, we define polarization states of all massive
vector states with respect to the common momentum $p$ with $p^2=m^2$. For 
a state with momentum $(p-q)$ with $(p-q)^2=m^2$ we have
\begin{align}
\varepsilon_{v\mu}(p-q)=\Lambda_{\mu}^{~\nu}(p-q,p)\,\varepsilon_{v\nu}(p)\,,
\end{align}
using the Lorentz-boost $\Lambda^{\mu}_{~\nu}$,
\begin{equation}
\Lambda^\mu_{~\nu}(p-q,p) = \delta^\mu_{~\nu}  + \frac{1}{q^2-4m^2}
\left[2\frac{q^2}{m^2}(p-q)^\mu\, p_\nu + 4q^\mu\,p_\nu - 4\left(p-\frac{q}{2}\right)^\mu q_\nu\right]\;,
\label{eqn:boost}
\end{equation}
which gives $(p-q)^\mu=\Lambda^\mu_{~\nu}\, p^\nu$. Note that the little-group
indices $v$ of the polarization vector $\varepsilon_{v\mu}(p)$ are unchanged.
Outer products of vector-boson states associated to the same momentum are
transverse tensors and can be expressed in terms of a basis of spin operators 
(\ref{eqn:tensorBasis}), 
likewise the physical transverse part of the amplitude. 
A tensor decomposition of a spin amplitude $M^{v'v}$ is then
\begin{multline}
M^{v'v}=M^\mu_{~\rho} \,\Lambda_{\mu}^{~\nu}(p-q,p)\,\bar\varepsilon_{v'\nu}(p)\, \varepsilon_{v}^{\rho}(p)\\
 \rightarrow \quad 
M^\mu_{~\rho}\, \Lambda_{\mu}^{~\nu}(p-q,p) = 
	c\,  \mathbb{P}^\nu_{~\rho} + 
	c_\lambda (\mathbb{S}^\lambda)^\nu_{~\rho} 
	+ c_{\lambda\kappa} (\mathbb{S}^{\{\lambda}  \mathbb{S}^{\kappa\}})^\nu_{~\rho}  \;,
\end{multline}
In order to obtain this decomposition, we will first introduce an auxiliary
tensor basis,  which we relate to classical spin vectors in a second step.

The basis is constructed with three transverse vectors
\begin{align}
    v_i\, \in \, \{ \mathbb{P}\cdot q_1, \mathbb{P}\cdot q_2\,, \mathbb{P}\cdot u_2 \} \,,
\end{align}
where we project three vectors which are linear independent of $p$.
We use the following rank-two tensor basis,
\begin{equation}
\begin{split}
 \big\{ T_1^{\alpha\beta}, \ldots, T_9^{\alpha\beta} \big\} = 
	\Big\{\mathbb{P}^{\alpha\beta}\;, 
  	v_1^{[\alpha} v_2^{\beta]}\;,
  	&v_1^{[\alpha} v_3^{\beta]}\;,
  	v_2^{[\alpha} v_3^{\beta]}\;,  \\
  	& v_1^{\{\alpha} v_2^{\beta\}}\;,
  	v_2^{\{\alpha} v_3^{\beta\}}\;,
  	v_1^{\{\alpha} v_3^{\beta\}}\;,
  	v_1^{\{\alpha} v_1^{\beta\}}\;,
  	v_2^{\{\alpha} v_2^{\beta\}}\;
	\Big\}\,.
\label{eqn:opBasis}
\end{split}
\end{equation}
where we use the convention
\begin{align}
v_i^{[\alpha}v_j^{\beta]} &= \frac{1}{2}\left[ v_i^\alpha v_j^\beta - v_i^\beta v_j^\alpha\right]\;, \\
v_i^{\{\alpha} v_j^{\beta\}} &= \frac{1}{2} \left[ 
	v_i^\alpha v_j^\beta + v_i^\beta v_j^\alpha \right] 
	- \frac{1}{3}(v_i\cdot v_j) \,\mathbb{P}^{\alpha\beta} \;.
\end{align}
When contracted with explicit polarization states we obtain
\begin{align}
T_n^{v'v} = \bar\varepsilon_{v'}(p-q)\cdot T_n \cdot \varepsilon_{v}(p) 
	= \bar\varepsilon_{v'}(p)\cdot\left[\Lambda(p,p-q)\cdot T_n \right]\cdot \varepsilon_{v}(p) \,.
\end{align}
Next we relate the tensor basis to functions in classical spin. One method is
to use a Clebsch-Gordan decomposition of a product of polarization
states~\cite{Cangemi:2022abk},
\begin{equation}\label{eqn:polDec} \overline{\varepsilon}_{v'}^{\mu}(p)
\varepsilon_{v}^{\nu}(p) = \frac{1}{3}
\overline{\varepsilon}_{v'}\cdot\mathbb{P}\cdot\varepsilon_{v}\,\left(\eta^{\mu\nu}-\frac{p^\mu
p^\nu}{m^2} \right)- \frac{i}{2m}\epsilon^{\mu \nu \rho \sigma}p_{\rho}
\overline{\varepsilon}_{v'}\cdot\mathbb{S}_{\sigma}\cdot\varepsilon_{v}\ +
\overline{\varepsilon}_{v'}\cdot\mathbb{S}^{\{\mu}\mathbb{S}^{\nu\}}
\cdot\varepsilon_{v} \,, 
\end{equation}
which expresses the spin states in terms of a
irreducible  scalar, anti-symmetric and symmetric-traceless representation of
the little group transformations. 
Motivated by the relation
\begin{equation}
\frac{\overline{\varepsilon}_{v}\cdot\mathbb{S}^{\mu}\cdot\varepsilon_{v}}{\overline{\varepsilon}_{v}\cdot\varepsilon_{v}} 
\to S^{\mu}
\end{equation}
we obtain a simple replacement rule 
$T_n{} \rightarrow T_n^{\rm cl} $ 
which relates quantum-spin observables 
to classical ones, 
\begin{align}\label{eqn:clRepl}
T_n^{\rm cl}(S) = -\big[\Lambda(p, p-q)\cdot T_n\big]_{\mu\nu}
	\left[ \frac{1}{3}\left(\eta^{\mu\nu}-\frac{p^\mu p^\nu}{m^2} \right) 
	-\frac{i}{2m}\epsilon^{\mu \nu \rho \sigma}p_{\rho} S_{\sigma}
	+S^{\{\mu} S^{\nu\}}  \right]\,,
\end{align}
for the classical-spin vector $S^\mu$ with $S\cdot p=0$. Importantly we
introduce an overall minus sign in \cref{eqn:clRepl}, since we remove
inner product $\overline\epsilon_v\cdot\epsilon_v=-1$.\footnote{This sign
implies that the scalar component of the scalar-vector waveform matches the
scalar-scattering waveform.}

%--------------------------------------------------------------------------------
\subsection{Graviton States}
\label{sec:gravitonFF}
%--------------------------------------------------------------------------------

In order to make the classical limit manifest, we use a form-factor
decomposition for the graviton states.
The graviton polarisation tensors are $\varepsilon_{\mu\nu}$, which we write as
product states of two spin-1 fields. Using a basis of plus and minus polarized
vectors, the polarization tensors of a circularly polarized gravitational wave
are given by
\begin{align}\label{eqn:helicityStates}
    \varepsilon_{\pm 2}^{\mu\nu}(k) = \varepsilon_{\pm1}^{\mu}(k) \varepsilon_{\pm1}^{\nu}(k).
\end{align}
and hence fulfill the transversality condition $\varepsilon_{\pm 2}^{\mu\nu}(k)\,
k_\mu = \varepsilon_{\pm 2}^{\mu\nu}(k)\, k_\nu = 0$. The states are normalised
as
\begin{align} \label{eqn:NormalisationPolTensors}
\bar\varepsilon_{s\mu\nu}\varepsilon_{s'}^{\mu\nu}= \delta_{s's}\,, 
\quad \varepsilon^{\mu\nu}_s = \bar\varepsilon^{\,\mu\nu}_{-s}\,, \quad s,s'\in \{ -2,2 \} \,,
\end{align}
and are traceless, e.g. ${({\varepsilon_s})_{\mu}}^{\mu}={({\bar\varepsilon_s})_{\mu}}^{\mu}=0$.
In analytic computations, contractions of the graviton states with the
linearly independent set of four vectors $\{u_1,u_2,k,q_1\}$ appear. As we are using product states, these factorize into two 
scalar products of a spin-1 polarization vector $\varepsilon_{\pm1}^{\mu}(k)$ with one of the independent momenta. 
Starting from 10 symmetric contractions, we can reduce to a minimal basis of only two quadratic monomials. 
First, we use the traceless condition which can be implemented by setting
$(\varepsilon_s)^2=0$ together with the Gram determinant identity $\mathrm{G}(u_1, u_2, k, q_1,
\varepsilon_h)=0$ to arrive at 9 independent monomials.
Taking into account also transversality $\varepsilon_h\cdot k=0$, this set is reduced by 4
monomials. Next, we use the gauge freedom $\varepsilon_h\rightarrow
\varepsilon_h+k$ to set inner products $\varepsilon_h\cdot u_1$ to zero,
\begin{align}\label{eqn:PolGauge}
    u_1\cdot \varepsilon_h = 0
\end{align}
We arrive at the fact that amplitudes including a single graviton emission can
be given as linear combinations of the two monomials
\begin{align}\label{eqn:gravitonMon}
F_1^{2h}=(\varepsilon_{h}\cdot u_2) \, (\varepsilon_{h}\cdot u_2)\,,\qquad 
F_2^{2h}=(\varepsilon_{h}\cdot u_2) \, (\varepsilon_{h} \cdot q_1) \,.
\end{align}
This representation does not limit the generality of our result. If one wishes
to consider polarization vectors $\varepsilon_{h}$, the gauge condition can easily be imposed by sending
\begin{align}\label{eqn:unGauge}
\varepsilon_{h}  \rightarrow  \tilde\varepsilon_h=\varepsilon_{h} -\frac{u_1 \cdot \varepsilon_{h}}{u_1\cdot k} k \,,
\end{align}
and evaluating the amplitude using $\tilde\varepsilon_s$. In this way, full
generality can be restored.

%--------------------------------------------------------------------------------
\section{Waveform Observable from Amplitudes}
\label{sec:overview}
%--------------------------------------------------------------------------------
The goal of this work is to compute the waveform emitted during the scattering
process of two black holes, where one is rotating, at order $\mathcal{O}(G^3)$
including linear-in-spin effects.  For the waveform observable, this is explained
through the KMOC formalism \cite{Kosower:2018adc,Maybee:2019jus,Cristofoli:2021vyo}, which relates the waveform
to matrix elements including the S-matrix.
(See also the recent generalization of
such asymptotic observables in ref.~\cite{Caron-Huot:2023vxl}.)

In the following, we briefly summarize the necessary results to highlight the
connection between the gravitational waveform and scattering amplitudes. For
more detailed reviews we refer the reader to
refs.~\cite{Cristofoli:2021vyo,Herderschee:2023fxh,Elkhidir:2023dco}.
Let us start by focusing on the momentum-space waveform observable. It is given
by the matrix element ${\cal M}^{\vec s}$ (and its conjugate)
\begin{align}  \label{eq:M}
    i\hat{\delta}^D(q_1+q_2-k) \mathcal{M}^{\vec s}(p_i,q_i,k) & = 
   \langle {p_1'}^{s_1'} {p_2'}  |\, S^\dagger a^{s}(k)  S\, |p_1^{s_1} p_2 \rangle\,,
\end{align}
which depends on the spin and helicity quantum numbers of the particles
involved in the scattering process, which we collectively denote by $\vec
s=\{s_1,s_1',s\}$. Furthermore, we absorb some explicit factors of $2\pi$ by
defining
\begin{equation}
 \hdelta^D(x) \equiv (2\pi)^D\, \delta^D(x)\;.
\end{equation}
In order to make contact with the classical scattering of spinning point-like
objects, we introduce classical spin variables through a form-factor
decomposition \cite{FebresCordero:2022jts},
\begin{align}
{\cal M}^{\vec s}&=\sum_{i=1}^2\sum_{j=1}^4  {\cal M}^{ij}  \, 
	F_i^s \, T_j^{s_1's_1}\,,
\end{align}
using the operator basis (\ref{eqn:opBasis}).

The classical spin-dependence of the amplitude $\mathcal{M}^{\vec{s}}$ is
obtained by replacing the Pauli-Lubanski spin operator $\mathbb{S}^\mu$ by its
classical counterpart $S^\mu$.  In terms of the form factor decomposition, the
operators $T_n$ are replaced by $T_j^\cla(S)$ using \cref{eqn:clRepl},
\begin{align}
{\cal M}^{\cla,s}( S)=  
	\sum_{i=1}^2\sum_{n=1}^4 {\cal M}^{ij}\, F_i^s\, T^{\cla}_j(S)\Big|_{\rm class}   \,,
\end{align}
where we take the classical scaling limit of the building blocks of the matrix element
${\cal M}^{\vec s}$ (see \cref{sec:amplitude}).
This replacement relies on an observed match between the form-factors of
spin-$s$ theories and point-particle theories with classical spin to order $2s$, see
ref. \cite{Vaidya:2014kza}.  Commonly this relation is coined the \textit{spin
universality principle}, which states that the form-factors of spin operators
are universal and can be matched between different finite-in spin theories and
point particles with continuous spin. For the case at hand, this was verified 
to \G3 and $\mathcal{O}(S^2)$ \cite{FebresCordero:2022jts}.

We will also consider the spin-less system (\ref{eqn:matterLagrangianScalarI})
for which the form-factor decomposition only concerns the graviton state.
Alternatively, we obtain a matching waveform in the spin system: since the
matrix elements are obtained as a polynomial in the spin vector we recover
the spin-less waveform for $S^\mu=0$.

The position-space waveform in the classical limit (\ref{eqn:classLimit}) 
is given by the Fourier transformation \cite{Kosower:2018adc,Cristofoli:2021vyo}.
In fact, for large distances from the
scattering event, the metric field $h_{\mu\nu}(r)$ admits an
expansion in negative powers of $|\mathbf{r}|$. The waveform is identified as
the coefficient of the leading-order contribution in this expansion, e.g.
\begin{align}\label{eqn:waveform}
h(S,r,p_i,b)\big|_{|\mathbf{r}| \rightarrow\infty} = \frac{1}{4\pi |\mathbf{r}|} h^{\infty}(t-|\mathbf{r}|,S,\mathbf{n},p_i,b)\,.
\end{align}
Plugging in $k = \omega \hat k  = \omega (1,\mathbf{n})$, with $\mathbf{n}=\mathbf{r}/|\mathbf{r}|$, for the graviton momentum and taking
the Fourier transform to time-domain and impact parameter space, we have~\cite{Cristofoli:2021vyo}
\begin{align}\label{eqn:FT}
    h^{\infty}(u,S,\mathbf{n},p_i,b)&= \frac{i\kappa}{2}\int_0^\infty 
	\frac{d\omega}{2\pi} 
	\int d^D\mu\,  
	e^{-i \omega u + i b_1\cdot q_1+i b_2\cdot q_2 }  
       \mathcal{M}^{{\cla},-2}(S,p_i,q_i,k) + \mbox{c.c.}\,,
\end{align}
where we define the retarded time $u = t-|\mathbf{r}|$ and introduced the shorthand notation
\begin{equation}
    d^D\mu=\left[\prod_{i=1}^2 \frac{d^D q_i}{(2\pi)^{D}}\hat{\delta}\big(2 \bm_i\, u_i\cdot q_i\big) \right]
            \hat{\delta}^{D}(q_1+q_2-k) \,.
\end{equation}
We split the impact-parameter vectors into symmetric and anti-symmetric 
contributions 
\begin{align}
b_1=b+b_s \,,\quad b_2=-b+b_s \;,
\end{align}
and omit the symmetric one ($b_s=0$), since it corresponds to a translation of the system, 
which can be retrieved from a shift in the time variable $t\rightarrow (u - b_s \cdot \hat k)$  
in the final observable.

In more conventional notation, the amplitude $h^{\infty}(u,S,{\bf n})$ is
related to \textit{plus} and \textit{cross} polarizations of the waveform,
which is commonly used in the data analysis of gravitational wave detectors,
\begin{align}
 h_{+} = {\rm Re}\,  \left[ h^{\infty}  + g_0 \right] \,, \qquad 
 h_{\times } = {\rm Im}\,  \left[ h^{\infty}  + g_0 \right]\,,
\end{align}
where $g_0$ accounts for a constant background of the metric that has to be
fixed through initial conditions.

\begin{figure}[t]
\centering
\includegraphics{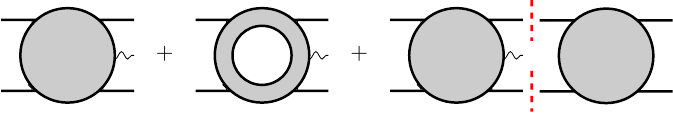}
\caption{The waveform amplitude is a combination of scattering amplitudes. Up
to one loop we require a tree amplitudes, a loop amplitude (the amplitude
contribution) and a phase-space integral over a product of tree amplitudes (the
cut contribution). We suppress disconnected contributions which do not
contribute for generic momenta and frequency.}
\label{fig:SaS}
\end{figure}

Let us now turn the discussion towards the explicit scattering amplitudes
necessary in the computation of the waveform.  As \cref{eq:M} is already
proportional to $\kappa$ we will need amplitudes up to $\mathcal{O}(\kappa^5)$.
To this end we express the S-matrix in terms of the transition matrix $T$, i.e.
using $S=1 + iT$. Plugging in this expansion, we find two contributions to
the amplitude \cite{Kosower:2018adc},
\begin{align}  \label{eq:Mexpanded}
 \langle {p_1'}^{s_1'} p_2' | S^\dagger a^s(k) S |p_1^{s_1} p_2\rangle & =
 i \langle {p_1'}^{s_1'} p_2' | a^s(k) T           |p_1^{s_1} p_2 \rangle
 + \langle {p_1'}^{s_1'} p_2' | T^\dagger a^s(k) T |p_1^{s_1} p_2\rangle \;,
\end{align}
with the first one being linear in $T$ and second one quadratic in $T$.  Below
we refer to the first term of \cref{eq:Mexpanded} as the {\it amplitude
contribution} and the second we denote the {\it cut contribution}. 
The amplitude contribution corresponds to the five-point scattering amplitude $M^{\vec s}$,
\begin{equation}
\langle {p_1'}^{s_1'} p_2' | a^s(k) T |p_1^{s_1} p_2 \rangle = \hat{\delta}^D(p_1'+p_2'+k-p_1-p_2) M^{\vec s}  \;,
\end{equation}
which we need at the tree and one-loop level,
\begin{align}
M^{\vec s} &= M^{{\rm tree},\vec s} + M^{1-{\rm loop},\vec s}  + {\cal O}(\kappa^6) \,.
\end{align}
These contributions are graphically represented by the first two terms of
\cref{fig:SaS}.  The second term, the cut contribution $C^{\vec s}$ is given by
\begin{equation}
\langle {p_1'}^{s_1'} p_2'  | T^\dagger a^s(k) T |p_1^{s_1} p_2\rangle = 
 i\hat{\delta}^D(p_1'+p_2'+k-p_1-p_2) C^{\vec s}  \;,
\end{equation}
which is turned into a
phase-space integral of a product of scattering amplitudes by inserting a
complete set of states,
\begin{multline}
\langle {p_1'}^{s_1'} p_2' | T^\dagger a^s(k) T |p_1^{s_1} p_2\rangle \\ = 
 \sum_{n} \int d{\rm LIPS}_n(\vec{\ell}\,)~ 
 \langle {p_1'}^{s_1'} p_2' | T^\dagger | \psi_n^{\vec p}(\vec{\ell}\,) \rangle~ 
 \langle \psi_n^{-\vec p}(-\vec{\ell}\,)  |    a^s(k) T |p_1^{s_1} p_2\rangle \,,
\end{multline}
where the sum runs over all n-particle scattering states and the phase-space
integration is performed over the respective multi-particle Lorentz-invariant
phase space (LIPS). The collection of phase-space momenta and particle
polarization labels are denoted by $\vec \ell=\{\ell_1,...,\ell_n\}$ and $\vec
p$, respectively. 

\begin{figure}[t] 
\begin{subfigure}[b]{0.48\linewidth}
 \centering
 \raisebox{0.5cm}{\includegraphics[scale=1.1]{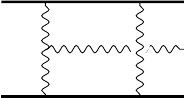}}
\end{subfigure}
\begin{subfigure}[b]{0.48\linewidth}
 \centering
 \includegraphics{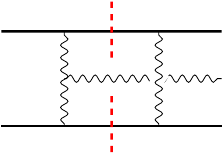}
\end{subfigure}
\caption{A one-loop pentagon integral and its two-particle cut. Both, the
Feynman integral and the phase-space integral, appear combined in the classical
waveform observable.}
\label{fig:cut}
\end{figure}

The sum over intermediate states simplifies in fixed-order computations. The phase-space integral does
not contribute at leading order (for $n=0,1$), because of particle flavor
conservation, as the two matter lines are associated to distinct particle
types.  At next-to-leading order, the insertion of two-particle states
contribute with two on-shell massive propagators.  This contribution corresponds
to a two-particle cut and is given by the product of a four-point and a
five-point tree,
\begin{multline}
\langle {p_1'}^{s_1'} p_2'  | T^\dagger a^s(k) T |p_1^{s_1} p_2\rangle = 
\sum_{\hat s_1}\int d\Phi_1(\ell_1)\, d\Phi_2(\ell_2)\, \delta^D(\ell_1+\ell_2 + p_1'+p_2')  \\
 \times \langle {p_1'}^{s_1'} p_2'| T^\dagger | \ell_1^{-\hat s_1} \ell_2 \rangle ~ 
\langle  (-\ell_1)^{\hat s_1} (-\ell_2)   |    a^s(k) T |p_1^{s_1} p_2\rangle  + {\cal O}(\kappa^6)\,,
\end{multline}
Here two massive lines appear, that are associated to a vector field
and the scalar. This contribution is depicted on the right of
\cref{fig:SaS}.  The corresponding phase-space measures are denoted by
$d\Phi_i$ and are defined in \cref{eqn:phaseSpaceMass}, while $\hat{s}_1$ is
the polarization label of the intermediate vector-field.  Additionally, there
could be contributions coming from disconnected diagrams including three-point
trees and propagators.  Since we are working with real-valued kinematics, these
pieces will only contribute at zero frequency of the emitted graviton. Such
a contribution can also be fixed through initial conditions, we do
not consider it in this work. Finally, we note that the four-point tree is
written in terms of $T^\dagger$ and is therefore evaluated with propagators
with the opposite $i\delta$ prescription 
\begin{equation}
 \frac{1}{p^2-m^2+i\delta}\quad \rightarrow \quad
 \frac{1}{p^2-m^2-i\delta}\;,
\end{equation}
as compared to the five-point tree and one-loop amplitudes.  We can write the
cut contribution as the Cutkosky cut of the one-loop (quantum) amplitude $M^{1-{\rm
loop},\vec s}$ in the $(p_1'+p_2')$-channel, as depicted in \cref{fig:cut}, 
\begin{align}\label{eqn:C}
C^{\vec s} &\equiv {\rm Cut}_{1'2'} \left[ M^{1-{\rm loop},\vec s} \, \right]\,.
\end{align}
The cut contribution plays an important role, by setting causality properties
of the total observable \cite{Caron-Huot:2023vxl}. Technically, the
cut terms also subtract hyper-classical contributions  from the observable
\cite{Brandhuber:2023hhy,Georgoudis:2023lgf,Herderschee:2023fxh}.
With this we have linked the asymptotic observables (\ref{eqn:FT}) directly to the
following scattering data
\begin{align}\label{eqn:waveamp}
{\cal M}^{\vec s} = M^{{\rm tree}, \vec s} + M^{1-{\rm loop},\vec s} + C^{\vec s} + {\cal O}(\kappa^6) \,.
\end{align}
In the next section, we discuss the evaluation of the amplitude functions in
the classical limit.

%--------------------------------------------------------------------------------
\section{Scattering Amplitudes Computation}
\label{sec:amplitude}
%--------------------------------------------------------------------------------

We now discuss the computation of the classical scattering amplitudes (\ref{eqn:waveamp}).
In order to leverage the organizational principle of QFT and
exploit established field-theory methods, we perform the computation in two
steps: we first compute the quantum scattering amplitudes and then consider
their classical limits (\ref{eqn:classLimit}).

An important aspect of our computation is that we base it on an exact numerical
approach.  For the computation of quantum scattering amplitudes  we use the
program \Caravel{} \cite{Abreu:2020xvt}. At tree level, the program yields
numerical results for scattering amplitudes in exact modular arithmetic.  At
loop level, the program implements the numerical unitarity method
\cite{Ossola:2006us,Ellis:2007br,Berger:2008sj,Giele:2008ve,Ellis:2008ir} in
the variant of refs.~\cite{Ita:2015tya,Abreu:2017xsl,Abreu:2017hqn}, which
provides numerical values for the rational integral coefficients of a one-loop
integral basis.
We employ functional reconstruction techniques
\cite{Peraro:2016wsq,vonManteuffel:2014ixa} to obtain  analytic expressions for
the classical expansions.  In order to deal with the classical limit of
helicity states, we compute the amplitudes in a form-factor decomposition.

%--------------------------------------------------------------------------------
\subsection{Organization of the Computation}
%--------------------------------------------------------------------------------
\begin{figure}
\begin{subfigure}{0.24\linewidth}
\includegraphics{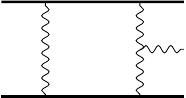}
\caption*{A}
\end{subfigure}
\begin{subfigure}{0.24\linewidth}
\includegraphics{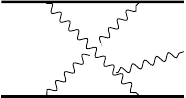}
\caption*{B}
\end{subfigure}
\begin{subfigure}{0.24\linewidth}
\includegraphics{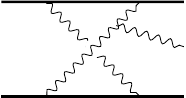}
\caption*{C}
\end{subfigure}
\begin{subfigure}{0.24\linewidth}
\includegraphics{figs/topology_D}
\caption*{D}
\end{subfigure}
\caption{Example Feynman diagrams for the scattering of massive particles
interacting through gravitational exchange. Here, wavy lines represent
massless gravitons, while solid lines refer to massive scalar or
vector-fields.}
\label{fig:topologies}
\end{figure}

For the computation of the waveform observable we proceed in two steps, we
first compute the one-loop amplitude, which admits the standard decomposition,
\begin{align}\label{eqn:intDec}
{M}^{1-{\rm loop},\vec s} = \sum_{\Gamma\in \Delta} c^{\vec s}_{\Gamma}(\epsilon)\, {\cal I}_{\Gamma} \,,
\end{align}
where $\Delta$ is the set of all one-loop diagrams. Furthermore,
$\mathcal{I}_\Gamma$ are the one-loop master integrals and
$c^{\vec{s}}_\Gamma(\epsilon)$ the corresponding $\epsilon$ dependent integral
coefficients.  
The set of master integrals is a subset of the set of
diagrams and we take this into account by allowing vanishing for integral
coefficients.

Next we compute the cut in the $(p_1'+p_2')$-channel by recycling the one-loop amplitude
computation.  The cut of the quantum amplitude in \cref{eqn:C}, is given as a
sum of phase-space integrals which is related to cut master-integrals,
\begin{align}\label{eqn:intDecCut}
{C}^{\vec s}  & = {\rm Cut}_{1'2'} \left[ \sum_{\Gamma\in \Delta} c^{\vec s}_{\Gamma}(\epsilon)\, \cI_{\Gamma}  \right] 
= \sum_{\Gamma\in \Delta } c^{\vec s}_{\Gamma}(\epsilon)\,  {\rm Cut}_{1'2'}\left[ \cI_{\Gamma} \right] \,, 
\end{align}
where we keep in mind that some of the cuts of the master integrals vanish.
The coefficients $c^{\vec s}_{\Gamma}(\epsilon)$ are not affected
by cutting and are those of \cref{eqn:intDec}.  Given the similarities between
the amplitude and the cut contribution we can combine both via 
\begin{align}\label{eqn:expDec}
{M}^{1-{\rm loop},\vec s} +  {C}^{\vec s} & = \sum_{\Gamma\in \Delta } c^{\vec s}_{\Gamma}(\epsilon)\, {\cI}^{\rm exp}_{\Gamma} \,, 
\end{align}
with the new type of master integrals given by
\begin{align}\label{eqn:Iexp}
{\cI}^{\rm exp}_{\Gamma}  & \equiv {\cI}_{\Gamma} +  {\rm Cut}_{1'2'} \left[ {\cI}_{\Gamma} \right ]\,.
\end{align}
We note that the integrals ${\cI}^{\rm exp}_{\Gamma}$ coincide with
${\cI}_{\Gamma}$ if the cut contribution in the $(p_1'+p_2')$-channel vanishes. 
In the classical limit, the integrals $\cI_\Gamma$ were computed in
refs.~\cite{Brandhuber:2023hhy,Georgoudis:2023lgf,Herderschee:2023fxh} and
their generalizations $\cI^{\rm exp}_\Gamma$ in ref.~\cite{Caron-Huot:2023vxl}.
We have validated the ${\cI}^{\textrm{exp}}_\Gamma$ integrals and collected our
independent computation in \cref{sec:FI}.

Finally, we remark that the cutting procedure of a fixed integral is
related to its discontinuity and imaginary part, however, this does
not hold true for the classical limit of the full amplitude. In the classical
limit, distinct cuts may not be properly resolved which leads to over counting,
such that a cut in the $(p_1+p_2)$-channel is indistinguishable from the one in
the $(p_1'+p_2')$-channel, since the difference between the momenta
$k=p_1+p_2-(p_1'+p_2')$ is considered to be a sub-leading contribution.
Working with a diagrammatic representation of the integrals the association of
quantum and classical cuts is manifest and this issue can be avoided.

In summary, we require two types of object, the integrals ${\cal I}^{\rm
exp}_{\Gamma}$ and the coefficients $c^{\vec s}_{\Gamma}(\epsilon)$, both
expanded in the classical limit. In the following section
\ref{sec:numUnitarity}, we discuss the explicit computation of these
ingredients within the framework of Numerical Unitarity.

%--------------------------------------------------------------------------------
\subsection{Brief Review of Numerical Unitarity}
\label{sec:numUnitarity}
%--------------------------------------------------------------------------------
For the computation of the amplitudes $M^{\vec s}$ we now compute the integral
coefficients, $c^{\vec s}_{\Gamma}(\epsilon)$ applying the numerical unitarity method
\cite{Ossola:2006us,Ellis:2007br,Berger:2008sj,Giele:2008ve,Ellis:2008ir}
 in the variant of ref.~\cite{Ita:2015tya,Abreu:2017xsl,Abreu:2017hqn}.  First, we promote the
integral decomposition \cref{eqn:intDec} to an integrand decomposition,
\begin{align}\label{eqn:integrandDec}
M^{\vec s}(\ell) = \sum_{\Gamma \in \Delta} 
	\sum_{k\in M_\Gamma \cup S_\Gamma } c^{\vec s}_{\Gamma,i}(\epsilon)\, 
	\frac{m_{\Gamma,i}(\ell)}{\prod_{j\in P_\Gamma} \rho_j(\ell) }
\end{align}
where $\rho_j$ denotes the propagator variable. $P_\Gamma$ denotes the set of
propagators associated to diagram $\Gamma$.  As opposed to the integral
decomposition formula in \cref{eqn:intDec}, the sum in \cref{eqn:integrandDec}
runs over integral insertions $m_{\Gamma,i}$ associated to master integrals
$M_{\Gamma}$ and surface terms $S_{\Gamma}$ of a given diagram $\Gamma$.  While
surface terms are necessary to parameterize the integrand, they will ultimately
integrate to zero.  We construct the surface terms from unitarity compatible
integration-by-parts identities
\cite{Ita:2015tya,Gluza:2010ws,Schabinger:2011dz,Larsen:2015ped}.

The coefficients $c_{\Gamma,i}$ are now obtained by exploiting the
factorization property of the loop integrand (\ref{eqn:integrandDec}).
Specifically, we consider loop-momentum values $\ell^\Gamma$ where the
propagators are on-shell, that is $\rho_j(\ell^\Gamma) = 0$ iff $j\in
P_\Gamma$. The leading contribution of the integrand in this limit behaves as
\begin{equation}\label{eq:onshell}
  \sum_{\vec s_i\in \text{states}}\prod_{i\in T_\Gamma}{M}_i^{{\rm tree},\vec s_i^{\,\prime}}(\ell^\Gamma)=
  \sum_{\Gamma'\geq\Gamma,i\in M_{\Gamma'}\cup S_{\Gamma'}}
 c^{\vec s}_{\Gamma',i}(\epsilon)\,  \frac{m_{\Gamma',i}(\ell^\Gamma)}
  {\prod_{j\in(P_{\Gamma'}\setminus P_\Gamma)}\rho_j(\ell^\Gamma)}\ .
\end{equation}
On the left-hand side of this equation, we denote by $T_\Gamma$ the set of tree
amplitudes associated with the vertices in the diagram corresponding to
$\Gamma$, and the sum is over the states propagating through the internal lines
of $\Gamma$.  On the right-hand side, we sum over integral topologies, denoted
by $\Gamma'$, which contribute to the limit, for which $P_\Gamma\subseteq
P_{\Gamma'}$.
The coefficients $c^{\vec s}_{\Gamma,i}(\epsilon)$ are obtained by solving the
system of linear equations given by \cref{eq:onshell} for a  sufficient number
of values of $\ell^\Gamma$. So far all step were performed numerically.  In
order to extract the dimensional dependence of the state sums, we vary the
dimension $D_s$ of the particle representations over four values, $D_s \in
\{5,6,7,8\}$ and extract the coefficients $c_n(\epsilon)$ of the following
parameterization of the integral coefficients,
\begin{equation}
 c_{\Gamma,i}^{\vec s}(D_s,\epsilon) = \frac{c_0(\epsilon)}{(D_s-2)^2} + \frac{c_1(\epsilon)}{(D_s-2)} + c_2(\epsilon) + c_3(\epsilon)\,(D_s-2)\;.
\end{equation}
The parameterization is set up to capture the explicit $D_s$ dependence of the
graviton propagators
\begin{equation}
 \frac{i}{p^2+i\delta}~\frac{1}{2}\left[
 \eta_{\mu\alpha}\,\eta_{\nu\beta} + \eta_{\mu\beta}\,\eta_{\nu\alpha} -
 \frac{2}{D_s-2}\eta_{\mu\nu}\,\eta_{\alpha\beta}\right]\;.
\end{equation}
Similarly, the dependence on the dimensional regulator $\epsilon$ is
reconstructed from sampling over finite numerical values of $\epsilon$. 
We evaluate the tree-level scattering amplitudes through Berends-Giele recursion
\cite{Berends:1987me} using an implementation of the Feynman rules obtained with the help of
\textsc{xAct}~\cite{Brizuela:2008ra,Nutma:2013zea}.
By performing all these calculations using finite-field arithmetic we are able to
determine the coefficients exactly as rational functions in $\epsilon$ and $D_s$ with 
no loss of numerical precision.
An important technical ingredient for the numerical amplitude computation is a rational
parameterization of phase space and the mass parameters. We provide the
technical details of this in \cref{sec:Computation}.

Having determined the $c^{\vec s}_{\Gamma,i}$ as functions in $\epsilon$ and
$D_s$ at a single numerical phase-space point we obtain the integrated
representation (\ref{eqn:intDec}) by dropping surface terms and replacing
master integrands, by master integrals. Given that every propagator structure
at one loop has at most one associated master integral we suppress the
corresponding label and use $c^{\vec s}_{\Gamma}$ instead of $c^{\vec s}_{\Gamma,1}$.

\begin{figure}
 \centering
 \includegraphics{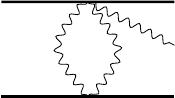}
 \hspace{2cm}
 \includegraphics{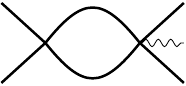}
 \hspace{2cm}
 \includegraphics{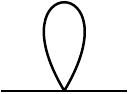}
\caption{An example of integrals that do not contribute in the classical limit.
Anticipating the cancellation the massless bubble contribution is dropped
already before we the classical limit. The massive bubble and tadpole integrals
turn scaleless and integrates to zero in dimensional regularization in the
classical limit.}
\label{fig:bubble}
\end{figure}

Finally, we remark, that we simplify the computation by omitting all tadpole
integrals and the bubble integral with two massless or two massive propagators.
In the unitarity computation, we omit the respective cuts in $\Delta$. The
diagrams are displayed in \cref{fig:bubble}.  Omitting these integrals early is
permitted, since they do not contribute in the classical limit, which we
anticipate.  We explicitly checked on random phase-space points that the bubble
diagram does in fact drop out in the classical limit.  All massive bubbles and
tadpoles reduce to scale-less integrals in the classical limit and vanish in
dimensional regularization. It is important to keep the pentagon integrals
although they only start contributing at higher orders in $\epsilon$.

%--------------------------------------------------------------------------------
\subsection{Form-factor Decomposition}
%--------------------------------------------------------------------------------
We now implement a form-factor decomposition of the graviton helicity states and the
polarization states of the massive vector field.  The motivation is two fold: 
\begin{enumerate}
\item Technically we need good control over the classical scaling limit  and
the analytic reconstruction, which is achieved by working with the rational
form factors instead of polarization states. The decomposition can also impact the compactness of the result.
We use gauge-symmetry and scaling considerations in the classical limit to
determine the form factors. 
\item We intend to relate scattering data of massive vector fields to a spin
variable. This step inherently relies on a form-factor decomposition in terms
of the Pauli-Lubanski operators introduced in sec.~\ref{sec:spinOperators}.
\end{enumerate}

In our effective field theory, the gravitational wave corresponds to the
expectation value of an external graviton and the polarization of the observed
wave is related to the polarization tensors $\varepsilon_{\mu\nu}$ of this
graviton.
For the scattering of four scalar fields and a graviton, we decompose the
amplitude in terms of
\begin{equation}
 O^s_i = F_i^s \,, \quad i=1,2\;.
\end{equation}
In the presence of vector bosons we instead choose the basis of spin and
helicity amplitudes by
\begin{align}\label{eqn:Os}
O^{\vec s}_{ij}=F_i^s T_j^{s_1's_1} \,, \quad i=1,2\,,\quad j=1,\ldots,9\,,
\end{align}
and their classical counterparts,
\begin{align}
O^{\cla,\vec{s}}_{ij}=F_i^s T_j^\cla \,, \quad i=1,2\,,\quad j=1,\ldots,9\,,
\end{align}
using the definition of the graviton monomials (\ref{eqn:gravitonMon}) and the
spin decomposition (\ref{eqn:clRepl}). 
We combine them as in \cref{eqn:Os} and consider observables in the
decomposition,
\begin{align}
{\cal M}^{\vec s}= \sum_n {\cal M}^{n}\,  O_{n}^{\vec s} \,,
\end{align}
where we use the multi-index notation $n=(ij)$ with $i=1,2$ and $j=1,9$.

For the rational tree amplitudes we have the helicity independent form 
factors $M^{{\rm tree}, n}$ according to
\begin{align}
M^{{\rm tree}, \vec s}  &= \sum_n M^{{\rm tree}, n}\, O_{n}^{\vec s} \,.
\end{align}
By construction, the number of independent spin and helicity configurations,
denoted by N=2 (N=18), of the tree amplitude $M^{\textrm{tree},\vec{s}}$
matches the number of independent tensors $O_n^{\vec{s}}$ of the scalar
\eqref{eqn:procScalar} and the scalar-vector \eqref{eqn:procVector} theory,
respectively.  Consequently, we extract the scalar form factors directly from 2
(18) helicity amplitudes by solving the linear system
\begin{equation}
\begin{pmatrix} 
 M^{\textrm{tree},\vec{s}_1} \\ \vdots \\ M^{\textrm{tree},\vec{s}_N} 
\end{pmatrix} = \begin{pmatrix}
 O_1^{\vec{s}_1} & \cdots & O_N^{\vec{s}_1} \\ 
 \vdots & \ddots & \vdots \\ 
 O_1^{\vec{s}_N} & \cdots & O_N^{\vec{s}_N} \
\end{pmatrix}\begin{pmatrix}
 M^{\textrm{tree},1} \\ \vdots \\ M^{\textrm{tree},N} 
\end{pmatrix}\;.
\label{eqn:tensor_decomposition}
\end{equation}
For the corresponding one-loop amplitudes, we obtain the form factors by
decomposing the integral coefficients
\begin{align}
c^{\vec s}_{\Gamma}(\epsilon)  &= \sum_n c^{n}_{\Gamma}(\epsilon)\,  O_{n}^{\vec s}\,,
\end{align}
such that the one-loop amplitude is given by,
\begin{align}
M^{1-{\rm loop}, n}  &= \sum_{\Gamma\in \Delta} c^{n}_{\Gamma}(\epsilon)\, {\cal I}_{\Gamma} \,.
\end{align}

%--------------------------------------------------------------------------------
\subsection{Classical Limit of Quantum Amplitudes}
%--------------------------------------------------------------------------------
To realize the classical limit in the quantum scattering amplitudes we follow
the procedure outlined below.  Due to the different analytical structure of
tree-level and one-loop amplitudes we proceed with slightly different
strategies. 

Our first goal is the analytic reconstruction in the scaling parameter $\bm$,
in order to perform the classical expansion.  For tree-level amplitudes the
classical contributions are given at the order $\bm^4$ (\ref{eqn:scalingMs}).
Using \cref{eqn:tensor_decomposition} allows to obtain the form factors
$M^{\textrm{tree},n}$ for a fixed momentum configuration as
\begin{equation}
 M^{\textrm{tree},n} = \sum_{k=-1}^{1} M^{{\rm tree},n}_k(D_s-2)^k\,.
\end{equation}
The coefficients $M^{{\rm tree},n}_k $ are rational functions in $\bm$ and we parameterize them as,
\begin{align}
M^{{\rm tree},n}_k(\bm)=\frac{1+\sum_{i=1}^{e_n} \bm^i N^n_{ki}}{ \sum_{j=1}^{\tilde e_n} \bm^j D^n_{kj} },
\end{align}
and compute the parameters $N_{ki}^n$ and $D_{ki}^n$ and the maximal polynomial
degrees $e_n$ and $\tilde e_n$, from a sufficient number of numerical evaluations of
$M^{{\rm tree},n}_k$ using Thiele's formula
\cite{abramowitz1964handbook,Peraro:2016wsq}.  We then obtain the classical
tree amplitude  as the leading term in the large-$\bm$ expansion,
\begin{align}
M^{{\rm tree},n}\Big|_{\rm class}=
	\bm^4 \sum_{k=-1}^{1} \frac{N^n_{ke_n}}{D^n_{k\tilde e_n}} (D_s-2)^k \,.
\end{align}
As dictated by the limit we kept the highest-order $\bm$-monomials in the
numerator and denominator of the rational function $M^{{\rm tree},n}_k(\bm)$.
At this point the ratio $({N^n_{ke_n}}/{D^n_{k\tilde e_n}})$ is a number and we
will reconstruct its analytic form using the techniques discussed in the next
section.

For the computation it is helpful to understand how the overall mass scaling
appears.  The graviton form factors $F_i$, given in \cref{eqn:gravitonMon}, are
homogeneous in $\bm$, so that the mass scaling is manifest in the coefficients
$M^{{\rm tree},n}_k$.
In contrast, the vector-scalar amplitude typically has spurious contributions
with $\bm^n$ for $n> 4$, which cancel once the amplitude is combined with the
classical form factors containing the spin vectors $S^\mu$.
E.g. the proper scaling is easily broken, if the boost factor (\ref{eqn:boost})
were to be omitted.  We have already discussed (see \cref{sec:vectorStates})
that the subspace transverse to the momentum $p$, in which the spin is defined,
plays a distinguished role in mapping polarization states to a classical spin
vector.
We attribute the cancellation of unphysical mass scaling to the alignment of
frames of the polarization vectors and the spin-tensor decomposition transverse
to the frame momentum $p$.
After the expansion of form factors as well as the tensor basis in the
classical limit the tree amplitude assumes the expected $\bm^4$ scaling.

For the one-loop integral coefficients we follow analogous steps to the above
and obtain a generalized power series in $\bm$ with rational dependence on
$D_s$ and $\epsilon$, 
\begin{align}\label{eqn:coeffExp}
c^{n}_{\Gamma}(\epsilon,D_s,\bm) = 
	\bm^{p^n_{\Gamma}}  
	\sum_{k=-2}^{1} 
	\sum_{l=0}^\infty 
	c^{n}_{\Gamma,kl}(\epsilon)
	(D_s-2)^k \frac{1}{\bm^l}  \,.
\end{align}
with a leading power $p^n_{\Gamma}$ in $\bm$ that depends on the integral
topology.

We now turn to the computation of the master integrals in the classical limit.
For convenience we split the integrals in soft and hard contributions in an
expansion by regions~\cite{Beneke:1997zp,Smirnov:1998vk},
\begin{align}\label{eq:EbR}
\lim_{\bm \rightarrow \infty} {\cal I}_{\Gamma} =  
\cI_{\Gamma}\big|_{\rm eik} + 
\cI_{\Gamma}\big|_{\rm hard} \,.
\end{align}
The integrals in the hard region are polynomial in the momentum-transfer
variables $q_i^2$ and do not contribute to the long-range observables; they
drop out in the Fourier transformation (see \cref{sec:FTUV}). We thus focus on
the eikonal integrals.

\begin{figure}[t]
\centering
\includegraphics{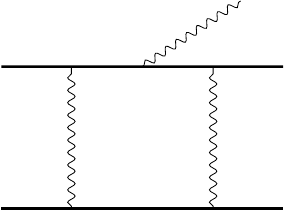}
\caption{Example pentagon diagram that that reduces via partial fractioning to other integral families.}
\label{fig:topologyPF}
\end{figure}
The expansion of the scalar integral basis leads to tensor integrals, which we
reduce into a basis of eikonal integrals using
\LiteRed~\cite{Lee:2012cn,Lee:2013mka} and partial fractioning of linearly
dependent propagators. An example integral that requires partial fractioning is
displayed in \cref{fig:topologyPF}.  We obtain the expansion,
\begin{align}\label{eqn:intExp}
{\cal I}^{\rm exp}_{\Gamma}\big|_{\rm eikonal}&=  \sum_{\gamma \in \Delta_{\rm eik}} d_{\Gamma,\gamma}(\epsilon)\, {\cal I}^{\rm eik}_{\gamma} \,,
\qquad
d_{\Gamma,\gamma} = \bm^{r_{\Gamma,\gamma}} \sum_{l=0}^\infty \frac{1}{\bm^l}\, d^l_{\Gamma,\gamma} (\epsilon)\,.
\end{align}
where only finitely many expansion terms in $\bar m$ are required. Here ${\cal
I}^{\rm eik}_\gamma$ is a master-integral basis of eikonal integrals.
$\Delta^{\rm eik}$ denotes the diagrams associated to the integral basis of
eikonal integrals. $r_{\Gamma,\gamma}$ gives the leading scaling of the scalar
master integral ${\cal I}^{\rm exp}_\Gamma$ in the classical limit.
Combining integrals and coefficients in the $\bm$ expansion, we obtain,
\begin{align}
\left[M^{1-{\rm loop},n}+C^{\vec n} \right]\Big|_{\rm class}= 
\bm^5 \sum_{k=-2}^{1} \sum_{\gamma\in\Delta^{\rm eik}} 
	(D_s-2)^k\,  
	c^{{\rm eik},n}_{\gamma,k}(\epsilon)
	\cI^{\rm eik}_\gamma \,,
\end{align}
where the coefficients $c^{{\rm eik},n}_{\gamma,k}$ are suitable linear
combinations of the expansions of the integral coefficients
(\ref{eqn:coeffExp}) and the integral reduction (\ref{eqn:intExp}). The
individual terms, the one-loop amplitude and the cut contributions have a
hyper-classical scaling with $\bm^6$, which drops out in the combination of the
terms
\cite{Brandhuber:2023hhy,Herderschee:2023fxh,Georgoudis:2023lgf,Caron-Huot:2023vxl}.

Finally we substitute an expansion of the eikonal master integrals in terms of
a functions basis $f_i$ (see \cref{sec:FI} for the definitions) and obtain the
classical limit of amplitude and cut contributions
\begin{align}\label{eqn:waveformFB}
\left[M^{1-{\rm loop},n}+C^{n} \right]\Big|_{\rm class} &= \sum_{i}  \left(\frac{r^n_{{\rm div},i}}{\epsilon} +r_i^n\right) f_i +{\cal O}(\epsilon)\,,\\
M^{{\rm tree},n}\Big|_{\rm class} &= r^n_{\rm tree} +{\cal O}(\epsilon) \,. 
\end{align}
We have used the 't Hooft-Veltman variant of dimensional regularization with
$D_s=4-2\epsilon$. Up to now we have discussed how to obtain numerical values
for the coefficients $r_i$ for a given set of external moments. Next we will
determine their analytic form from sampling over phase-space configurations.

%--------------------------------------------------------------------------------
\subsection{Analytic Reconstruction}
\label{sec:reconstruction}
%--------------------------------------------------------------------------------
We now discuss the computation of the analytic expressions for the 
rational functions $r^n_{\rm tree}$, $r^n_{{\rm div},i}$ and $r^n_i$ of
\cref{eqn:waveformFB}, by the functional-reconstruction method
\cite{Peraro:2016wsq,Abreu:2017xsl} in finite-field arithmetic
\cite{vonManteuffel:2014ixa}.

We start by evaluating the quantum scattering amplitudes on fixed momentum
configurations according to \cref{eqn:momenta}.
We generate several sets of external
momenta, for every set only a single kinematic invariant of
$\{y,\bm_1,\bm_2,\omega_1,\omega_2,q_1^2,q_2^2\}$ is varied. From these
univariate functions we can construct a naive ansatz for the rational
multivariate functions $\{r^n_{\rm tree}, r^n_{{\rm div},i}, r_i^n\}$ by simply
considering the minimal and maximal degree in each variable for the numerator
and denominator.
This ansatz can be simplified as each multivariate numerator and denominator
polynomial has to be homogeneous in the mass dimension. The mass dimensions can
be obtained by reconstructing the dependence on $t$ of an univariate slice
through all dimensionful parameters, such that
\begin{equation}
 (\bm_1,\bm_2,\omega_1,\omega_2) = \vec{a}_1 + \vec{a}_2\,t\;, \qquad (q_1^2,q_2^2) = \vec{a}_3 + \vec{a}_4\,t + \vec{a}_5\,t^2\;, \qquad y = a_6
 \label{eqn:slice_t}
\end{equation}
where $a_i$ are random numbers. The mass dimension is then given by the maximal
degree of $t$ for numerators and denominators. After these steps we now have a
rational parameterization of each of the functions $r^n_{\rm tree}$, $r^n_{\rm
div}$ and $r^n_i$.

We can further simplify the parameterization using the information of the $t$
dependence.  The rational coefficients of the pure integral functions $f_i$, as
well as the tree assume the form~\cite{Abreu:2018zmy},
\begin{align}\label{eqn:coeffForm}
r_i^n = \frac{ {\cal N}^n_i}{\prod_j w_j^{q^n_{ij}}}
\end{align}
where the denominator factors $w_i$ are given by the letters in the symbol
alphabet of the integral functions
\cite{Brandhuber:2023hhy,Herderschee:2023fxh,Georgoudis:2023lgf,Caron-Huot:2023vxl}
(see \cref{sec:FI}) with integer exponents $q_{ij}^n$. 
For simplicity we focus on the discussion on the coefficients $r^n_i$, however,
an analogous formula holds for $r_{\rm tree}^n$ and $r_{\rm div}^n$.
The numerators ${\cal N}^n_i$ are polynomials in the Lorentz invariants
(\ref{eqn:variables}). The letters that appear in the denominators are
collected in \cref{eqn:denFactors}.

Knowing the letters $w_i$ we obtain their functional form $w_i(t)$ by
evaluating them for the kinematics given in \cref{eqn:slice_t}. Using
univariate factorization we can then determine the denominator exponents
$q_{ij}^n$.

All that remains is to determine the numerator polynomials, for which we
write a polynomial parameterization.  We then set up a system of linear
equations for all unknown coefficients and generate sufficient numerical
samples of the classical expanded scattering amplitude to obtain solutions. In
this way we determine the numerator polynomial with coefficients in a finite
field.  Finally, we observe that we need two different prime numbers to lift
the functional reconstruction of the coefficient functions from the
finite-field to the rational numbers.  With this we have obtained the analytic
form of all functions $r_{\rm tree}^n$, $r_{{\rm div}}^n$ and $r_i^n$ for the
scalar-scalar and the scalar-vector spectral waveforms.

%--------------------------------------------------------------------------------
\section{Results}
\label{sec:results}
%--------------------------------------------------------------------------------
In this section we collect the classical waveform amplitudes
$\mathcal{M}^{\vec{s}}$, as given by \cref{eqn:waveamp}, in their leading order
and next-to-leading order approximations. Furthermore, we will discuss in depth the
analytical properties of the spectral waveforms that we have
observed.

%--------------------------------------------------------------------------------
\subsection{Leading-order Amplitudes}
\label{sec:tree}
%--------------------------------------------------------------------------------
The tree-level five-point amplitude for four massive scalar particles and a graviton
is given in the ancillary files. The amplitude is given in terms of a form-factor 
decomposition
\begin{equation}
 \mathcal{M}^{\textrm{tree},s} = \mathcal{M}_1\,F_1^s + \mathcal{M}_2\,F_2^s\;.
\end{equation}
The scattering amplitude is equivalent to the one obtained in ref. \cite{Luna:2017dtq},
which reads 
\begin{multline}
\mathcal{M}^{{\rm tree},s} =   -\frac{\kappa^3 \bar m_1^2 \bar m_2^2 }{4} \,
	\varepsilon_{s,\mu\nu}\left[
		\frac{4 P^\mu  P^\nu}{q_1^2 q_2^2} + \frac{2 y}{q_1^2 q_2^2} \big(Q^\mu P^\nu + P^\mu Q^\nu\big) \right. \\
	 \left.+ \left(y^2 - \frac{1}{D_s-2}\right) \left(\frac{Q^\mu Q^\nu}{q_1^2 q_2^2} - \frac{ P^\mu P^\nu}{\omega_1^2 \omega_2^2} \right)\right]\,,
\end{multline}
with
\begin{equation}
 P^\mu = -\omega_1 u_2^\mu + \omega_2 u_1^\mu\,, \qquad 
 Q^\mu = (q_1-q_2)^\mu  + \frac{q_1^2}{\omega_1} u_1^\mu - \frac{q_2^2}{\omega_2 } u_2^\mu \,,
\end{equation}
where $D_s$ is the dimensional regularization parameter associated to the
spin-degrees of freedom. Furthermore, once linear-in-spin effects are taken
into account the scalar-vector tree-level amplitude reads
\begin{align}
\mathcal{M}^{{\rm tree},s}(S) = \mathcal{M}^{{\rm tree},s} + \sum_{i=1}^2\sum_{j=2}^4 \mathcal{M}^{{\rm tree},ij}F^s_i T^{\rm cl}_j(S) + \mathcal{O}(S^2)\,.
\end{align}
We cross checked  the spin tree-amplitude against an independent computation
using the worldline quantum field theory formalism matching 
ref.~\cite{Jakobsen:2021smu,Jakobsen:2021lvp,Jakobsen:2023oow}.  The explicit results are
given in ancillary files. In particular, in the $S^\mu\rightarrow 0$ limit, we
recover the tree amplitude of the scalar-scalar scattering.

%--------------------------------------------------------------------------------
\subsection{Next-to-leading Order Waveform}
\label{sec:loop}
%--------------------------------------------------------------------------------
At the one-loop level we divide the waveform amplitude into its tree-level part
and a part that is infrared (IR) divergent, one that is ultraviolet (UV)
divergent, a tail, and finite contributions. Thus, the waveform amplitude reads
\begin{align}
\mathcal{M} =  
\mathcal{M}^{{\rm tree}}  
+ \mathcal{M}^{{\rm IR}} 
+  \mathcal{M}^{{\rm UV}}
+  \mathcal{M}^{{\rm tail}} + \mathcal{M}^{{\rm finite}}\,.
\end{align}
Let us comment briefly on the individual terms. The tree contribution is
identical to the one presented in \cref{sec:tree}. Next, the IR divergent terms
are a phase and can be absorbed into a redefinition of the coordinate time in
the position-space waveform as shown below. The term is explicitly given by
\begin{align}
\mathcal{M}^{{\rm IR}} &=
\left[ \frac{1}{\epsilon} - \log\left(\frac{\mu^2_{\rm IR}}{\mu^2}\right)\right] \mathcal{W}_S\, \mathcal{M}^{{\rm tree}}  \;.
\end{align}
The soft factor of the waveform $\mathcal{W}_S$ is discussed below in \cref{sec:IR}.
A derivation of $\mathcal{W}_S$ can be found in \cref{sec:Weinberg}.
Furthermore, the UV divergent contributions are given by
\begin{align}
 \mathcal{M}^{{\rm UV}} = \left[\frac{1}{\epsilon} - \log\left(\frac{\mu^2_{\rm UV}}{\mu^2}\right) \right] \overline{\mathcal{M}}^{{\rm UV}} \;,
\end{align}
where $\overline{\mathcal{M}}^{UV}$ is local in momentum-transfer variables
$q_1$ and $q_2$. Consequently they yield local contribution in impact parameter
space and drop out of the far-field waveform.
The tail and finite terms combine to give the characteristic waveform patterns
of gravitational scattering processes.  The tail contribution is given by
\begin{align}
 \mathcal{M}^{{\rm tail}} =
 -\log\left(\frac{\omega_1\omega_2}{{\mu}_{\rm IR}^2}\right) \mathcal{W}_S\,\mathcal{M}^{{\rm tree}}
- \log\left(\frac{\omega_1\omega_2}{{\mu}_{\rm UV}^2}\right) \, \overline{\mathcal{M}}^{{\rm UV}} \,,
\end{align}
while the finite remainder is defined through
\begin{align}
 \mathcal{M}^{{\rm finite}} = \sum_{i=1}^{18} f_i r_i  \,,
\end{align}
with the functions $f_i$ given in \cref{eqn:fbasisfin}. We provide
ancillary files containing the waveform observable at 
$\mathcal{O}(G^3)$ up to linear in spin corrections. The folder
structure is as follows:
\begin{enumerate}
    \item[] \texttt{anc/waveform.m}
    \item[] \texttt{anc/loadWaveform.wl}
\end{enumerate}
The file \texttt{waveform.m} contains the LO ($G^2$) and NLO ($G^3$)
waveforms with and without spin. The spin dependence is given in terms of
the form-factor decomposition in \cref{eqn:clRepl}. The notebook \texttt{loadWaveform.wl} 
contains the notation and commands to load and utilize the waveforms.

%--------------------------------------------------------------------------------
\subsubsection*{Waveform for Two Spinning Objects}
%--------------------------------------------------------------------------------
We have computed the scattering of a vector boson and a scalar and focus on
linear in spin corrections for one black hole. At this order, the spin corrections 
for the second black hole can be
obtained from symmetrization. However, some care is required due to the gauge
condition $u_1\cdot\varepsilon_h=0$.  A naive symmetrization would lead to
expressions with two gauge conventions. First one needs to remove this gauge
condition by the replacement $\varepsilon_h\rightarrow \tilde\varepsilon_h$
(\ref{eqn:unGauge}), after which symmetrization can be done consistently.

%--------------------------------------------------------------------------------
\subsubsection*{Infrared Singularities}
\label{sec:IR}
%--------------------------------------------------------------------------------
Both contributions, the amplitude and its cut, develop IR divergences.  In
ref.~\cite{Weinberg:1965nx} it has been shown that the IR pole factorizes into
a {\it soft factor} and the tree-level amplitude. The divergence is related to
soft gravitons being exchanged between all pairs of external particles.  The
same logic applies to the cut amplitude, but now only graviton
exchanges that are consistent with the cut topology in \cref{fig:cut} are taken into account. We
review the soft factors for the amplitude and the cut contributions in
\cref{sec:Weinberg}.  This leads to a subtraction of the super-classical
contributions in the soft factor of the complete waveform amplitude $\mathcal{M}$ and additionally alters the classical piece.
The combined IR pole is given by \cite{Caron-Huot:2023vxl},
\begin{align}
    \mathcal{M}\big|_{\mathrm{IR}} &=  \mathcal{W}_S\,\mathcal{M}^{\mathrm{tree}}\;, \quad \text{with} \quad
	\mathcal{W}_S=  i G(\bm_1 \omega_1 + \bm_2 \omega_2)
		\frac{1}{\epsilon}\left(1 + \frac{y(2y^2-3)}{2(y^2-1)^{3/2}}  \right)\,.
\end{align}
Here we suppress spin and helicity labels, since the same relation holds for
the scalar-scalar as well as the vector-scalar scattering process.

The result exponentiates into a total phase and can safely be
absorbed into an infinite redefinition of the retarded time $u$,
more precisely,
\begin{multline}
	e^{-i\omega u}\mathcal{M}  = e^{-i\omega \Big [u -\left( \frac{1}{\epsilon} -  \log\frac{\mu_{\rm IR}^2}{\mu^2} \right)\frac{\mathcal{W}_S}{\omega}\Big]}
	\left(M^{\mathrm{tree}}   +  \mathcal{M}^{{\rm UV}}
	+  \mathcal{M}^{{\rm tail}} + \mathcal{M}^{{\rm finite}}\right) +\mathcal{O}(G^{3})\,.
\end{multline}
In fact, this redefinition has an interpretation in classical general
relativity. The term,
\begin{equation}
    \Delta u_{\rm out} = \frac{2 G}{\omega} (\bar m_1 \omega_1 + \bar m_2 \omega_2)\log\left( \frac{r_\mathrm{obs}}{r_{\mathrm{min}}}\right)\,,
\end{equation}
corresponds to the Shapiro delay~\cite{Shapiro:1964uw}, where
$r_{\mathrm{min}}$ is the distance at closest approach and $r_{\mathrm{obs}}$
is the distance of the observer.
It is interpreted as the time delay a graviton
experiences when escaping the gravitational potential of the massive bodies, as
was already pointed out in ref.~\cite{Brandhuber:2023hhy}. The cut contribution
gives rise to an additional time-delay (or advance), 
\begin{equation}
    \Delta u_{\rm in} = \frac{2 G}{\omega} (\bm_1 \omega_1 + \bm_2 \omega_2) \frac{y(2y^2-3)}{2(y^2-1)^{3/2}}\log\left( \frac{r_\mathrm{in}}{r_{\mathrm{min}}}\right)\,.
\end{equation}
where $r_\mathrm{in}$ is the initial distance between the two black holes.
This result was obtained in a classical analysis~\cite{Caron-Huot:2023vxl} by
integrating along a geodesic in Schwarzschild spacetime from infinity to closest
approach. It thus originates from the deflection of the trajectory of an incoming
particle due to general relativistic effects. We have verified that the classical
derivation also holds in a Kerr space time and that the time delay is equivalent
for spinning and non-spinning black holes, as is also predicted by
Weinberg's theorem. Geometrically, this is due to the suppression of the angular
momentum terms in the Kerr metric at large radial distance.

%--------------------------------------------------------------------------------
\subsubsection*{Ultraviolet Singularities}
\label{sec:FTUV}
%--------------------------------------------------------------------------------
%
\begin{figure}[t]
\centering
\includegraphics{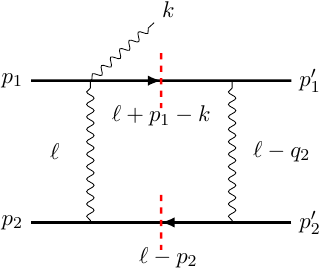}
\caption{A box cut that is part of the cut contribution. The momenta
$\{p_1,p_2\}$ are incoming, $\{p_1',p_2',k\}$ outgoing and $\ell$ flows
clockwise in the diagram. }
\label{fig:boxCut}
\end{figure}
Apart from the soft divergence predicted by Weinberg's theorem there is an
additional $\epsilon$ pole $\mathcal{M}^{\rm UV}$ coming from the cut of the
amplitude.  For the scalar expectation value it is proportional to,
\begin{equation}\label{eqn:UVexpr}
 \mathcal{M}\big|_{\mathrm{UV}} =
         -\left(\frac{\kappa}{2}\right)^5\frac{i }{8\pi\epsilon} (\bar{m}_1\omega_1 + \bar{m}_2\omega_2)  
         \frac{\bar{m}_1^2 \bar{m}_2^2 
	(\omega_1^2+\omega_2^2+ y\omega_1 \omega_2) 
	(1-2 y^2)^2}
	{\omega_1 \omega_2^3 (y^2-1)^{3/2}}~F_1^{2h}\,.
\end{equation}
The pole does not correspond to a UV divergence of the quantum amplitude but is
instead introduced by the eikonal expansion.
To discuss the origin of the additional UV divergence in the eikonal integrals,
we consider the box integral \cref{fig:boxCut},
\begin{align}\label{eqn:example}
\cI_{\rm box} =  \int \frac{d^D\ell}{(2\pi)^D} \frac{1}{\ell^2\, [(\ell+p_1-k)^2-m_1^2] (\ell-q_2)^2 [(\ell-p_2)^2-m_2^2]  } \,,
\end{align}
where it is understood that all propagators carry an additional $+i\delta$.
In the $\bar{m} \to \infty$ limit the integrals appearing in our calculation
are split into eikonal ($\ell\sim \bar{m}^0$) and hard region ($\ell\sim
\bar{m}^1$) integrals (see \cref{eq:EbR}).  The expansion of the second massive
propagator  in these regions is given by
\begin{equation}
\frac{1}{(\ell-p_2)^2-m_2^2+i\delta} = 
\begin{cases}
\frac{1}{-2\bar{m}_2 u_2\cdot \ell+i\delta}
-\frac{\ell\cdot(\ell- q_2)}{[-2\bar{m}_2 u_2\cdot \ell+i\delta]^2}+\dots &(\mathrm{eikonal})\,,\\
 \frac{1}{(\ell-\bar{m}_2 u_2)^2-\bar{m}_2^2+i\delta}
+\frac{\ell\cdot q_2}{[(\ell-\bar{m}_2 u_2)^2-\bar{m}_2^2+i\delta]^2}+\dots&(\mathrm{hard})\,.
\end{cases}
\end{equation} 
Similar expressions hold for the expansion of the second massive propagator.
The singular behavior of the expanded integrals is related to the original
unexpanded integral, but may differ in an important way.  While the eikonal
(hard) expansion matches the IR (UV) behavior of the unexpanded integral,
additional divergences develop in the UV (IR) of the eikonal (hard) expansion.
These additional divergences have to cancel between the eikonal and the hard
region since there is no corresponding divergence in the original integral.  

The appearance of UV singularities in the eikonal integral is seen from the
expansion of the massive propagators. The leading term in the eikonal expansion
scales as $1/\ell$ in the UV limit ($\ell\to\infty$), however, each additional
expansion order increases the scaling degree by one.  It follows that even if
an unexpanded integral containing massive propagators is UV finite, additional
UV divergences will be introduced at some order of its eikonal expansion.
Whether a UV divergence does in fact appear depends on the expansion order. For
the integral at hand we require an eikonal expansion to order $\bar{m}^{-4}$,
which is UV divergent.  

In our computation, UV contributions appear in intermediate stages of the
computation in the classical limit. They can be seen to cancel in the one-loop
amplitude, which produces a $1/\epsilon$ pole that matches the universal IR
factorization \cite{Weinberg:1965nx}.
In contrast, we do observe a non-vanishing UV contribution from the cut
contribution, defined in \cref{eqn:C}. We have verified this in two ways: 
\begin{enumerate}
\item We have compared the $1/\epsilon$ pole of the cut contribution to
Weinberg's soft theorem and found a mismatch. To this end we adjusted
Weinberg's theorem to the waveform computation (see \cref{sec:Weinberg}).
\item  We explicitly computed the UV pole of the cut. This was done by adding a
regulator mass $\mu_{\textrm{IR}}^2$ into the quadratic propagators before evaluating the cut
of the eikonal expansion in the UV limit.  For the above example $\cI_{D,{\rm
box}}$ (\ref{eqn:example}) this amounts to,
\begin{align}
\frac{1}{\ell^2-\mu_{\rm IR}^2+i\delta } \,,\qquad  \frac{1}{(\ell-q_2)^2 - \mu_{\rm IR}^2 + i \delta}\,.
\end{align}
The mass regulates the IR singularities, so that the full $1/\epsilon$ pole can
be associated to the UV. This computation produced the difference of the
$1/\epsilon$ pole of the cut contribution compared to Weinberg's infrared
theorem \cite{Weinberg:1965nx}.  
\end{enumerate}
Finally, we observe that the UV contribution has the expected analytic
properties. In fact the  UV contribution of the eikonal region lies in the
intersection of the hard and the eikonal functions; since the two must cancel.
It thus has analytic properties of the hard and the eikonal contributions, i.e.
it is polynomial in the momentum-transfer variables $q_i$. We observe this
property in \cref{eqn:UVexpr} and the analogous expression for the spin
waveform.  

We end this section with a remark about the phase space of the two-particle cut.
This phase space is compact for the quantum amplitude, namely a
two-dimensional sphere in four dimensions determined by the zero set of the
quadratic polynomials,
\begin{align}
(\ell+p_1-k)^2-m_1^2 =  (\ell-p_2)^2-m_2^2 =0 \,.
\end{align}
Consequently, there is no large-momentum region associated to this cut. This is
in contrast to the two-particle cut in the eikonal limit, where two eikonal
propagators are forced onto their mass shell,
\begin{align}
u_2\cdot \ell =  u_2\cdot \ell =0 \,.
\end{align}
These on-shell conditions are linear, such that
the resulting phase-space integral is performed over a two-dimensional plane,
which is non-compact. The boundary of this phase space is then a large momentum
region attributed to the UV.  This argument supports the surprising fact that
the cut contribution does yield a UV divergence.

%--------------------------------------------------------------------------------
\subsection{Analytic Properties}
%--------------------------------------------------------------------------------

The rational coefficients of the pure integral functions $f_i$, as well as the
tree take a particular form (\ref{eqn:denFactors}), which has interesting
physics implications.  We observe that the following letters 
\begin{equation}\label{eqn:denFactors}
\begin{split}
\{ w_i \}_{i=1,15}= 
\Big\{&\omega_1,\omega_2,q_1^2,q_2^2,y+1,y-1, 
\omega_2^2-q_1^2,
\omega_1^2-q_2^2, \\
&\omega_1^2+q_2^2(y^2-1),
\omega_2^2+q_1^2(y^2-1),
\omega_1^2-2 \omega_1 \omega_2 y+\omega_2^2,  \\
& 4 \omega_2^2 q_2^2+(q_1^2-q_2^2)^2,
4 \omega_1^2 q_1^2+(q_1^2-q_2^2)^2, \\
&\omega_1^2 q_1^4 - 2 \omega_1 \omega_2\, q_1^2 q_2^2 y+\omega_2^2 q_2^4,\\
& 4 \omega_1^2 (\omega_2^2 - q_1^2) + 4 \omega_1 \omega_2 y (q_1^2 + q_2^2) - 
 4 \omega_2^2 q_2^2 + (y^2 - 1) (q_1^2 - q_2^2)^2
\Big\}
\end{split}
\end{equation}
appear in the denominators, where $w_9,w_{10}, w_{14}$ and $w_{15}$ are absent
in the scalar waveform. 
Most of the $15$ letters are related to Gram determinants. 
We note that, in addition to the integral reduction, 
the form-factor decomposition introduces Gram determinants in the denominator.
The latter denominators cancel when transverse spin vectors and 
polarization states for the graviton are introduced.
To be concrete, we find that (up to overall numerical factors)
\begin{equation}
\{w_7,w_8\} \sim \Big\{ G(q_1,u_2),\, G(q_2,u_1) \Big\}\;,
\end{equation}
as well as
\begin{multline}
\{w_9,\ldots,w_{13}\} \sim \Big\{ G(u_1,u_2,q_2),\, G(u_1,u_2,q_1), \\
G(u_1,u_2,q_1+q_2),\, G(u_2,q_1,q_2),\, G(u_1,q_1,q_2)\Big\}\;,
\end{multline}
and finally
\begin{equation}
 w_{15} \sim G(u_1,u_2,q_1,q_2)\;.
\end{equation}
Note that $G(u_1,u_2,q_1,q_2)$ is equivalent to $\textrm{tr}_5^2$, where
\begin{equation}
 \textrm{tr}_5 = 4i\, \epsilon_{\mu\nu\alpha\beta}\, u_1^\mu u_2^\nu q_1^\alpha q_2^\beta\;.
\end{equation}
Therefore, the appearance of $w_{15}$ in the denominators corresponds directly
to a double pole.  At last, we want to remark that $w_{14}$ is related to the
modified Cayley determinant (see e.g.  ref.~\cite{Abreu:2017mtm}) of the
linearized pentagon integral.

The form (\ref{eqn:coeffForm}) has interesting physical implications: First of
all, the mass parameters do not appear in the function alphabet. Consequently
they appear in polynomial form through ${\cal N}_i^n$, as observed in
ref.~\cite{Vines:2018gqi,Damour:2019lcq}.   Second, the rational form of the
coefficients implies that the waveform at fixed-order is determined by a finite
set of polynomial coefficients of the numerator and denominator functions
${\cal N}_i^n$ and $w_i$.  Finally, in a further expansion, such as a PN
expansion, the expansion of the denominator letters imply repetitive patterns.
In light of that, only a finite number of terms is required to obtain the full PM
expansion, provided the integral functions and the function alphabet is known. 

In order to simplify our expressions, we exploit the observation that partial fractioned
coefficients $r_i^n$ take a simple form, with reduced mass dimension of
numerator and denominator. We use the algorithm \cite{Abreu:2019odu} (see also
\cite{Heller:2021qkz}) based on polynomial division to obtain compact analytic
forms of the waveform functions.

%--------------------------------------------------------------------------------
\section{Fourier Transformation}
\label{sec:fourier}
%--------------------------------------------------------------------------------
In order to make contact with a signal that could be observed in a
gravitational wave detector, we must perform the Fourier transformation to
time-domain and impact-parameter space given in \cref{eqn:FT}. After absorbing
the IR divergences into the redefinition of the retarded time, we are left with
the integral
\begin{align}\label{eq:FTspecific}
	h^{\infty} = \frac{i\kappa}{2} \int_0^\infty \frac{d\omega}{2\pi} \int d^D\mu\, e^{-i\omega u + i b\cdot q_1 }
	\left(\mathcal{M}_{\rm tree}^{-2}+ \mathcal{M}_{\rm finite}^{-2} + \mathcal{M}_{\rm tail}^{-2} + \mathcal{M}_{\rm UV}^{-2}+ {\rm c.c.} \right).
\end{align}
The UV divergent contribution is polynomial in $q_i$ and integrates to $\delta(|b|)$ in the Fourier
transformation, being at the same time associated to the hard region (\ref{eq:EbR}).
Since such terms do not contribute to the far-field waveform, we drop the UV
contribution from now on. 
The tree, the finite one-loop waveform
and the tail term yield finite results.

\begin{figure}[t]
	\centering
	\includegraphics[width = 0.4\textwidth]{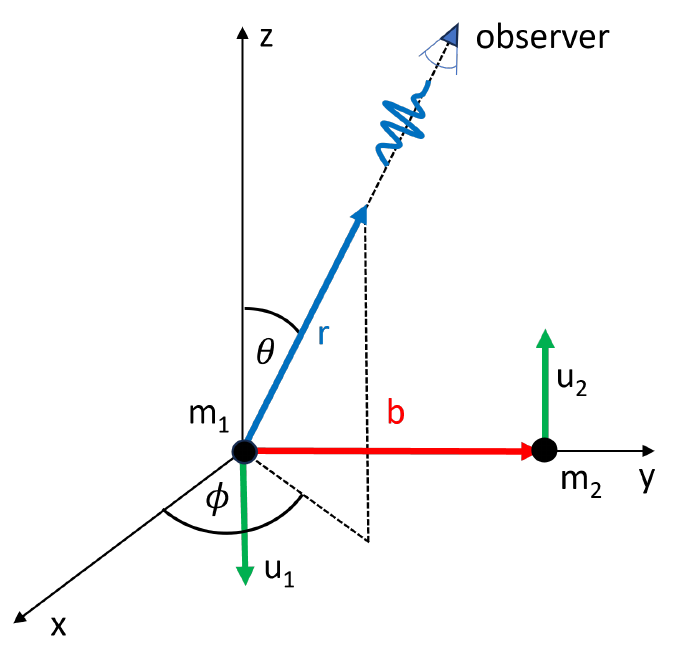}
	\caption{Parameters and coordinate system for observer of radiation from two-object scattering process.}
	\label{fig:observer}
\end{figure}

We start by defining a reference frame and initial conditions for the external
kinematics. We choose the frame shown in fig.~\ref{fig:observer} where the
position of the observer is given in terms of Euler angles. 
Hence, the graviton momentum and the $(-2)$-polarization are
\begin{align}
	k^{\mu}
	   = \omega \begin{pmatrix}
		1\\
		\cos{\phi}\sin{\theta}\\
		\sin{\phi}\sin{\theta}\\
		\cos{\theta}
		\end{pmatrix}\,,\quad
  \varepsilon_{-2}^{\mu\nu} = \varepsilon_{-}^{\mu}\varepsilon_{-}^{\nu} \quad \mathrm{with} \quad \varepsilon_{-}^\mu = \frac{1}{\sqrt{2}}\begin{pmatrix}
        0\\
        \cos \phi \cos \theta + i \sin \phi \\
        \cos \theta \sin\phi - i \cos\phi\\
        -\sin\theta
    \end{pmatrix}\,.
\end{align}
As suggested by ref.~\cite{Bini:2023fiz}, we define the external states and
their initial separation to be with respect to $u_1$ and $u_2$,
\begin{align}
	u_1^{\mu} = (
        y,\,  0,\,  0,\,  -\sqrt{y^2-1} )\,, \quad 
	u_2^{\mu} = (  y,\,  0,\,  0,\, \sqrt{y^2-1})\,, \quad 
	b^{\mu} = ( 0,\,|b|,\, 0,\,  0)\,.
\end{align}
Since $y$ corresponds to the Lorentz factor it is related to the initial
relative velocity $v$ of the two black holes through $y = 1/\sqrt{1-v^2}$. We will
show plots of waveforms for the observer at position $\phi = 7 \pi/10$, $\theta
= 7 \pi/5$ and a range of initial velocities. All plots are for an equal mass
black hole binary, i.e. $q=1$, with $m_1 = m_2 = 1 M_\odot$. Typical LIGO-Virgo-KAGRA
source waveforms with $m_i\sim 30 M_\odot$ would only differ by an overall
scale from what we present. Note that the difference between $m_i$ and $\bm_i$
is subsubleading in the classical expansion and we hence neglect it in the
Fourier transformation.
We now turn our attention to the actual integration in \cref{eq:FTspecific}.
First, $q_2$ is integrated out using the delta-function enforcing
energy-momentum conservation. Next, $q_1$ is parameterized by
\begin{equation}\label{eqn:q1Parametrization}
    q_1^{\mu} = z_1 u_1^{\mu} + z_2 u_2^{\mu} + z_b \frac{b^{\mu}}{|b|} + z_v \frac{v^\mu}{|v|} \, \quad \mathrm{with}\quad 
	v^\mu =\epsilon^{\mu\nu\rho\sigma} u_{1\nu} u_{2\rho}b_{\sigma} \,,
\end{equation}
which introduces an overall Jacobi-factor of $1/\sqrt{y^2-1}$. The integral over
$z_1$ and $z_2$ can be evaluated by solving the two constraints given in the
delta functions introduced by the on-shell conditions. The problem is hence
reduced to the integration over the three variables $z_b$, $z_v$ and $\omega$.
For these final steps, we follow the semi-analytical approach outlined in
ref.~\cite{Herderschee:2023fxh}. In the following, we focus on the
calculation of the different pieces in \cref{eq:FTspecific}. As a final point,
we note that there could be a non-dynamical background contributing to the
waveform, since we have been consistently neglecting zero frequency ($\omega =
0$) terms. Luckily, a gravitational wave detector only measures the changes in
the curvature of space-time. We hence `gauge' our signal by subtracting the
value at $u/|b| = -300$. 

\begin{figure}[t]
	\centering
	\includegraphics[width=\textwidth]{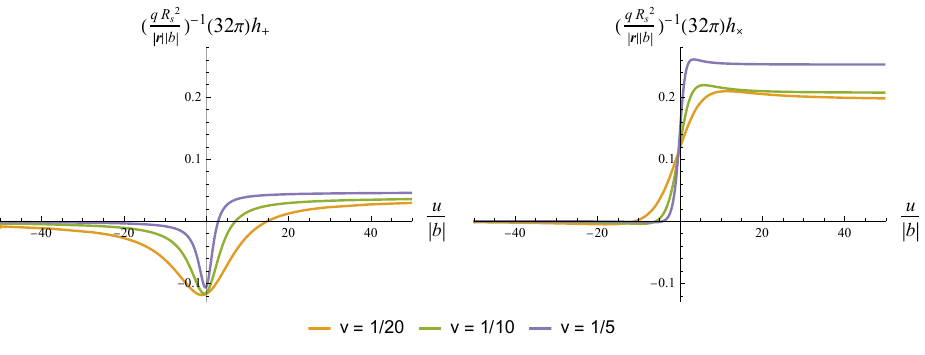}
	\caption{The leading-order waveform emitted during the scattering of two Schwarzschild black holes.}
	\label{fig:scalarTree}
\end{figure}

The Fourier transformation of the tree can be performed fully analytical with
the use of the residue theorem. Additionally, poles at infinity are regularized
using a principle value prescription~\cite{DeAngelis:2023lvf}. We show results
for two Schwarzschild black holes in fig.~\ref{fig:scalarTree} for an observer at $\phi = 7 \pi/10$ ,
$\theta = 7 \pi/5$.

\begin{figure}[b]
	\centering
	\includegraphics[width=\textwidth]{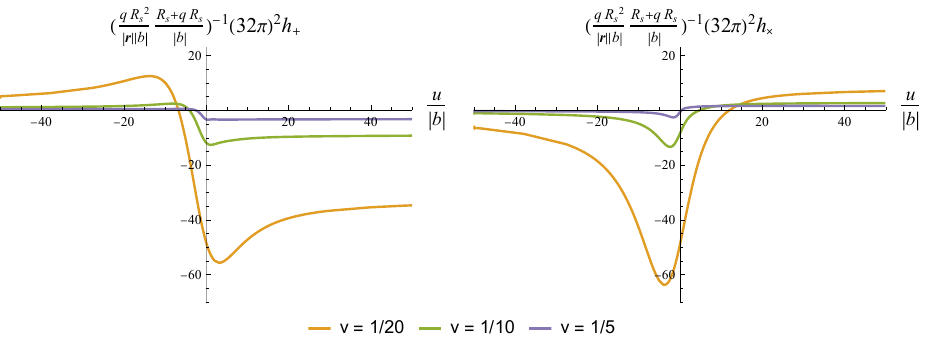}
	\caption{The next-to-leading order correction to the waveform emitted during the scattering of 
	two Schwarzschild black holes. The scale-dependent tail contribution is omitted.}
	\label{fig:nloFull}
\end{figure}

The one-loop waveform $\mathcal{M}^{\mathrm{finite}}$ contains functions, such
as square-roots and logarithms, not present at tree level, which complicates
the task significantly. As explained in detail in
ref.~\cite{Herderschee:2023fxh}, the integration over $\omega$ naturally splits
the waveform into two pieces, where one can be written as derivatives of a
principle value $\mathrm{PV}(\frac{1}{u/|b| + z_b})$ and the other as
derivatives of a delta function $\delta(u/|b| + z_b)$. This allows for an
analytic integration of $z_b$ using residues. The numerical integration over
$z_v$ is stable for the scalar waveform. When discarding the cut-contribution,
we reproduce fig.~(11) of ref.~\cite{Herderschee:2023fxh}. Adding the cut
amplitude significantly changes the waveform and results are displayed for a
range of velocities at position  $\phi = 7 \pi/10$, $\theta = 7 \pi/5$ in
fig.~\ref{fig:nloFull}. 
We note, that the choice made on how to separate the tail from the
finite piece significantly influences this result.

The spin corrected waveform introduces additional challenges 
for the stable numerical integration over $z_v$ after following the outlined steps. We
hence leave this Fourier transformation to future work.

%--------------------------------------------------------------------------------
\section{Validation}
%--------------------------------------------------------------------------------
Let us summarize in this section the cross checks that we have performed on our
computations.
\begin{description}
\item [Amplitude Computation:] The numerical computation of the integral
coefficients $c_{\Gamma,i}^{\vec{s}}(\epsilon)$ is performed using the
well-tested \Caravel{} framework. The necessary tree-level scattering
amplitudes involved in the unitarity cut computations have been already used at
the two-loop level \cite{FebresCordero:2022jts}, which yields strong
consistency checks of the amplitudes. For every phase-space point we perform
the so-called $N=N$ check, i.e.  we determine the all coefficients
$c_{\Gamma,i}^{\vec{s}}(\epsilon)$ by sampling \cref{eq:onshell} sufficiently
often. Afterwards, we generate an additional on-shell loop-momentum
configuration $\ell^\Gamma$ and compute both sides of \cref{eq:onshell} to
check if the integrand parameterization spans the entire amplitude.

\item [Master Integrals:] We computed independently all necessary master
integrals and their corresponding cut.  We extracted analytical expressions
from high-precision evaluations using \textsc{AMFlow} \cite{Liu:2022chg}, which
we validated against \pySecDec{} \cite{Borowka:2017idc}. Furthermore, we found
perfect agreement with ref. \cite{Caron-Huot:2023vxl} after translating
conventions.

\item [Analytical Reconstruction:] The analytically reconstructed amplitude
expressions are validated by numerical means. We evaluate them at a phase-space
point in a finite-field for a cardinality that was not used in the
reconstruction and compare the results with a direct numerical computation
within \Caravel{} using the same cardinality. Furthermore, we checked that the
classical expanded amplitude is independent of the choice of basis for the
intermediate form-factor decomposition by changing the tensors in
\cref{eqn:opBasis}.

\item [Spectral Waveform:] For tree-level scalar-scalar scattering
we reproduce the result of ref. \cite{Luna:2017dtq}, while the scalar-vector
tree-level amplitude is validated against an independent computation using the
worldline quantum field theory formalism. We also reproduce the results of ref. \cite{Jakobsen:2021lvp}.  The one-loop building block for the
scalar-scalar scattering has been validated against previous results of refs.
\cite{Brandhuber:2023hhy,Herderschee:2023fxh}. Furthermore, we also find
agreement with the independent computations \cite{Comparison,Brandhuber:2023} for the
cut contribution including the presence of additional UV divergences and 
with~\cite{Georgoudis:2023} up to $q_i$-independent terms and an overall factor $i$. Finally,
for both, scalar-scalar and scalar-vector, the infrared divergences of the
combined amplitude and cut contributions agree with Weinberg's infrared
factorization formula \cite{Weinberg:1965nx}. Finally, we verified that the
scalar-vector amplitude reduces to the scalar-scalar one in the limit $S \to
0$.

\item[Fourier Tranformation:] We confirm the cancellation of all unphysical
poles in the spectral waveform. We perform this check by analyzing the
vanishing locus of the denominator factors of \cref{eqn:denFactors} on the
physical phase space. That is, we evaluate the spectral waveform, see
\cref{eqn:FT}, for fixed helicity and spin assignment on generic lines through
the vanishing loci and numerically confirm that the functions are smooth.
Ignoring the cut contribution we also reproduce fig. (11) of ref.
\cite{Herderschee:2023fxh}.
\end{description}

%--------------------------------------------------------------------------------
\section{Conclusion}
\label{sec:conclusions}
%--------------------------------------------------------------------------------
We computed the gravitational Bremsstrahlung to next-to-leading order, i.e at
$\mathcal{O}(G^3)$ in the gravitational coupling expansion in the weak field
regime. We also compute for the first time linear-in-spin ($S^\mu$) effects. To
formulate the classical observable we study the time dependent expectation
value of the metric fluctuation, which is interpreted as the waveform of the
emitted gravitational wave. To organize the computation we embed the system
into minimally coupled second-quantized field theories, which we consider in a
scaling limit, associated to the classical point-particle dynamics. The results
extend recent computations of the same observable
\cite{Brandhuber:2023hhy,Herderschee:2023fxh,Georgoudis:2023lgf}, including 
cut contributions \cite{Caron-Huot:2023vxl,Comparison,Brandhuber:2023,Georgoudis:2023} and spin terms.

We consider two types of theories. Two distinct massive scalar theories, which
yield the dynamics of spin-less black-hole scattering. To describe the
scattering of a spinning and a spin-less black hole, we use the minimally
coupled scalar theory, as well as minimally coupled massive vector fields (Proca
theory). In this setup one can obtain up to quadratic in spin corrections
\cite{Vaidya:2014kza,FebresCordero:2022jts}, and we focused here on the linear term.

The classical observable is obtained from tree-level and one-loop scattering 
amplitudes and its two-particle cut in the eikonal limit.
The results validate a generalization of the universal IR factorization formula of Weinberg 
and a recent prediction for the IR of the classical waveform observable
\cite{Caron-Huot:2023vxl}.  Surprisingly, we encounter as well a UV
divergence in intermediate steps. The UV divergence is associate to the overlap
of the eikonal and hard region of the Feynman integrals in their classical
limit. The divergence is local in the momentum transfer and drops out after
Fourier transformation to impact parameter space in the far field.
In this context we also point out another surprising feature of the classical limit, 
namely that UV singularities can be obtained in one-loop cuts.

We observe interesting analytic properties in the spectral waveform. The
coefficients of the transcendental and algebraic basis functions are rational
expressions, whose denominators are symbol letters of the functions
characteristic alphabet of the basis of integral functions.  We anticipate that this
property leads to repeated patterns in further expansions of the present
post-Minkowskian result, e.g. in the post-Newtonian limit. 

For the computation of scattering amplitudes, we applied the numerical
unitarity approach in combination with analytic reconstruction. In order to
find compact final expressions, we used partial-fraction decomposition based 
on methods in algebraic geometry. Within this setup we anticipate that 
higher-spin corrections are well within reach in the near future.
Finally, we are looking forward to further developments towards the computation
of higher-order gravitational corrections and the advancement of 
multi-loop integration techniques and the classical waveform observable. 

\section*{Acknowledgments}
We gratefully acknowledge collaboration with Fernando Febres Cordero on
the implementation of one-loop unitarity cuts in the \Caravel{} program.
We particularly thank Henrik Johansson and Lucile Cangemi for many discussions
and hospitality at NORDITA.
We would like to thank Andreas Brandhuber, Graham Brown, Gang Chen, Stefano De
Angelis, Joshua Gowdy, Gabriele Travaglini, Alessandro Georgoudis, Carlo
Heissenberg, Ingrid Vazquez-Holm for very helpful discussions and the
comparison of the scalar spectral waveform.  We are particularly grateful to
Radu Roiban and Fei Teng for discussions and their cross-checks for the scalar results.
We thank 
Samuel Abreu,
Vittorio Del Duca,
Paolo Di Vecchia,
Riccardo Gonzo, 
Sara G\"undogdu, 
Enrico Herrmann, 
Lucio Mayer,
Donal O'Connell,
Paolo Pichini,
Jan Plefka,
Rafael Porto,
Michael Ruf,
Rudolfo Russo, 
Chia-Hsien Shen,
Jan Steinhoff,
Mao Zeng 
for discussions.
LB acknowledges support from the Swiss National Science Foundation (SNSF) under
the grant 200020 192092. LB acknowledges NORDITA for generously hosting them
during the three-month Visiting PhD Fellow Program.
M.K. is supported by the DGAPA-PAPIIT grant IA102224 at UNAM.
The authors acknowledge the Instituto de F\'{i}sica (UNAM) for providing
computing infrastructure and Carlos Ernesto L\'{o}pez Natar\'{e}n for his HPC
support. 

\appendix
%--------------------------------------------------------------------------------
\section{Field Theory}
\label{sec:FieldTheoryApp}
%--------------------------------------------------------------------------------
The Einstein-Hilbert Lagrangian ${\cal L}_{\rm EH}$ is given by,
\begin{align}
{\cal L}_{\rm EH} = -\frac{2}{\kappa^2} \sqrt{|g|} R\,, 
\end{align}
where $g={\rm det}(g_{\mu\nu})$ and $R$ the Ricci scalar contraction
$R=R^{\rho}_{~\mu\rho\nu}\,g^{\mu\nu}$ of the Riemann tensor. We use the conventions,
\begin{align}
 R^{\mu}_{~\nu \rho \sigma} &=  
 \partial_{\rho} \Gamma^{\mu}_{~\nu \sigma}-\partial_{\sigma} \Gamma^{\mu}_{~\nu \rho}
 +\Gamma^{\mu}_{~\alpha \rho}\Gamma^{\alpha}_{~\nu\sigma}-\Gamma^{\mu}_{~\alpha \sigma}\Gamma^{\alpha}_{~\nu \rho} \,,\\
 \Gamma^\mu_{~\nu\rho} &= \frac{1}{2} g^{\mu\sigma}\left[
     \partial_\nu g_{\sigma \rho} 
     + \partial_\rho g_{\sigma \nu} 
     - \partial_\sigma g_{\nu\rho} \right].
 \end{align}
The linearized Riemann tensor is given by,
\begin{equation}\label{eqn:linRiemann}
  R_{\mu\nu\rho\sigma} = - \frac{\kappa}{2}\big[\partial_\mu\partial_\rho h_{\nu\sigma} - \partial_\mu\partial_\sigma h_{\nu\rho} - \partial_\nu\partial_\rho h_{\mu\sigma} + \partial_\nu\partial_\sigma h_{\mu\sigma} \big]\;.
\end{equation}

In order to clarify our conventions, we introduce quantum fields and state spaces next. 
The graviton operator can be expanded in its modes as 
\begin{equation}
\mathbb{h}_{\mu \nu}(r) =  \sum_{h=\pm2} \int d \Phi_k \big[a^h(k) \bar\varepsilon_{h\mu\nu}(k) e^{-i k\cdot r} +  a^{h\dagger}(k) \varepsilon_{h\mu\nu}(k) e^{i k\cdot r }  \big]
\end{equation} 
where we drop the unphysical modes for simplicity.
The polarization tensors for the graviton are defined in terms of products of
vector states as given by \cref{eqn:helicityStates}.
The measure factors and commutation conventions are,
\begin{align}
    d\Phi_k&= \frac{d^Dk}{(2\pi)^D}\, \theta(k^0)\hdelta(k^2)\,,\quad 
    \left[ a^h(k) , a^{h'\dagger}(k') \right]  = (2 E_k) \hdelta^{D-1}(k-k') \delta^{hh'}\,.
\end{align}
and vanishing other commutators and we suppress un-physical modes.

Analogously, the massive vector and scalar fields are described by
\begin{align}
\mathbb{V}^{\mu}(r) &=  \sum_{v=1}^3 \int d \Phi_p \big[b^v(p) \bar\varepsilon_v^{\mu}(p) e^{-i p\cdot r} +  b^{v\dagger}(p) \varepsilon_v^{\mu}(p) e^{i p\cdot r }  \big]\,,\\
\mathbb{\Phi}(r) &=  \int d \Phi_p \big[b(p) e^{-i p\cdot r} +  b^{\dagger}(p) e^{i p\cdot r }  \big]\,.
\end{align}
To simplify the notation we will use the symbols $b^s$ and $b^{s\dagger}$ to
denote vector states and scalar states, with the index $s$ labeling the
different modes,
\begin{equation}
\big\{ b^0(p),\, b^1(p),\,b^2(p)\,, b^3(p) \big\}  = \big\{ b(p),\, b^1(p),\,b^2(p),\, b^3(p) \big\}\,.
\end{equation}
The mass $m$ of the associated states is encoded by the square of the momentum
argument $p^2=m^2$.  Phase space measure and creation/annihilation operators
for matter the fields $\mathbb{\Phi}$ and $\mathbb{V}^\mu$ are,
\begin{align}\label{eqn:phaseSpaceMass}
   d\Phi_{p_i}&= \frac{d^Dp_i}{(2\pi)^D}\, \theta(p^0)\hdelta(p_i^2-m_i^2)\,,\quad 
\left[ b^{s_1}(k) , b^{s_2\dagger}(k') \right] = (2 E_k) \hdelta^{D-1}(k-k') \delta^{s_1 s_2}\,,
\end{align}
We use vector-boson polarisation states that are real and normalised,
\begin{align} 
\varepsilon_{v\mu}(p)\varepsilon_{v'}^{\mu}(p)&= -\delta_{v'v}\,, 
\end{align}
and 
\begin{align} 
\bar \varepsilon_{v\mu}(p) = \varepsilon_{v\mu}(p)\,,\quad
\bar\varepsilon_{v\mu}(p)\varepsilon_{v'\nu}(p) \delta^{vv'} = \frac{p_\mu p_\nu}{p^2} -\eta_{\mu\nu}\,.
\end{align}
To be explicit, we consider boosts of the configuration with momentum
$p=(m,0,0,0)$ and polarisation states pointing into $e_x,e_y$ and $e_z$. Given
that the boost are unique only up to little-group rotations, the basis of
states is not unique either.  
\iffalse
Below we will be careful to define observables in a way to remove this little-group variance. We will not require 
explicit formulas for these states.
We consider the quantum fields to create in states.
The general way to write the massive initial states is
\begin{align}
    |\psi_{\rm in}\rangle = 
    \sum_{s_1,s_2} \int d\Phi_{p_1} d\Phi_{p_2} \phi_1^{s_1}(p_1) \phi_2^{s_2}(p_2) e^{i (b_1\cdot p_1+b_2\cdot p_2)}|p_1^{s_1} p_2^{s_2}\rangle\,,   
\end{align}
where each wave function is normalize to unity as
\begin{equation}
    \sum_s\int d \Phi_p |\phi_{i}^s(p)|^2 = 1 
\end{equation}
and the states are defined as,
\begin{equation}
| k^h \rangle = a^{h\dagger} (k) |0 \rangle \,, \quad | p^s \rangle = b^{s\dagger} (p) |0 \rangle \,.
\end{equation}
Following our compact notation $| p^0 \rangle$ denotes a scalar state, while $| p^v \rangle,\,v=1,2,3$, are the vector-boson polarisation states.

\draftnote{Add the sectin of Little group here?}
\fi

%--------------------------------------------------------------------------------
\section{Feynman Integrals}
\label{sec:FI}
%--------------------------------------------------------------------------------
In this section, we review the set of Feynman integrals required for the
one-loop computation. We follow the ideas of ref.~\cite{Caron-Huot:2023vxl}.
Representative Feynman diagrams of the five-point scattering process are shown
in \cref{fig:topologies}. We consider these diagrams in the classical limit.
\begin{figure}[t]
\centering
\includegraphics{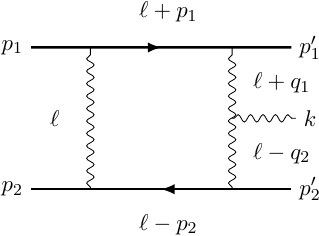}
\caption{Topology $A$ pentagon integral. The momenta $\{p_1,p_2\}$ are incoming, 
$\{p_1',p_2',k\}$ outgoing and $\ell$ flows clockwise in the diagram. }
\label{fig:topologyA}
\end{figure}
The pentagon integral family of topology $A$ in \cref{fig:topologies} before
the classical limit is given by
\begin{equation}
 {\cI}_A(\vec{\nu}) \equiv e^{\gamma_E\epsilon}
\int \frac{d^D\ell}{i\pi^{D/2}} \frac{1}{\rho_1^{\nu_1}\rho_2^{\nu_2}\rho_3^{\nu_3}\rho_4^{\nu_4}\rho_5^{\nu_5}}\;,
\end{equation}
with
\begin{align}
 &\rho_1 = \ell^2 +i\delta\;,  &&\rho_2 = (\ell + p_1)^2 - m_1^2 +i\delta\;, &&&\rho_3 = (\ell+q_1)^2 + i\delta\;, \\
 &\rho_4 = (\ell-q_2)^2 + i\delta\;, &&\rho_5 = (\ell- p_2)^2 - m_2^2 + i\delta\;. \nonumber
\end{align}
The vector $\vec{\nu} = (\nu_1,\nu_2,\nu_3,\nu_4,\nu_5)$ denotes the propagator
powers. We are only interested in the classical limit of these integrals.
Using the momentum parametrization in \cref{eqn:momenta} and taking the limit
$\bm \to \infty$ we obtain the following propagators,
\begin{align}\label{eqn:propCl}
 &\rho_1 = \ell^2 +i\delta\;, &&\rho_2 = 2(\ell\cdot u_1) +i\delta\;, &&&\rho_3 = (\ell+q_1)^2 + i\delta\;, \\
 &\rho_4 = (\ell-q_2)^2 + i\delta\;, &&\rho_5 = -2(\ell\cdot u_2) + i\delta\;. \nonumber
\end{align}
In the following, we denote the classical integral family simply by
${\cI}_A(\vec{\nu})$, where we understand that all propagators are taken in
their classical limit (see \cref{eqn:propCl}). For the classical integral
family we construct a pure basis of $16$ master integrals. First, we define the
following $6$ arguments of square roots,
\begin{align}
 &r_1 = -q_1^2\;, &&r_2 = -q_2^2\;, &&& r_3 = y^2-1\;, \\
 &r_4 = \omega_1^2 -q_2^2\;, &&r_5 = \omega_2^2 - q_1^2\;, &&& \Delta_5 = G(u_1,u_2,q_1,q_2)\;,
\end{align}
where we have defined the Gram determinant by
\begin{equation}
 G(\{p_i\}) = \det(\{p_j \cdot p_k\})\;.
\end{equation}
Our choice of the $16$ pure master integrals of $\cI_A$ read,
\begin{align}\label{eqn:pure}
 &M_1^A = \epsilon(1-2\epsilon)\, \cI_A[1,0,1,0,0]\;,  &&M_2^A = \epsilon(1-2\epsilon) \,\cI_A[1,0,0,1,0]\;, \nonumber \\
 &M_3^A = \frac{\epsilon(1-2\epsilon)}{\omega_1}\,\cI_A[0,1,0,1,0]\;, &&M_4^A= \frac{\epsilon(1-2\epsilon)}{\omega_2}\, \cI_A[0,0,1,0,1]\;, \nonumber \\
 &M_5^A = \epsilon^2\sqrt{r_1}\, \cI_A[1,1,1,0,0]\;, &&M_6^A = \epsilon^2\sqrt{r_4}\, \cI_A[1,1,0,1,0]\;, \nonumber \\
 &M_7^A = \epsilon^2\sqrt{r_5}\, \cI_A[1,0,1,0,1]\;, &&M_8^A = \epsilon^2\sqrt{r_3}\, \cI_A[0,1,1,0,1]\;, \\
 &M_9^A = \epsilon^2\sqrt{r_2}\, \cI_A[1,0,0,1,1]\;, &&M_{10}^A = \epsilon^2\sqrt{r_3}\, \cI_A[0,1,0,1,1]\;, \nonumber \\
 &M_{11}^A = \epsilon^2q_1^2\,\omega_1\, \cI_A[1,1,1,1,0]\;, &&M_{12}^A = \epsilon^2\sqrt{r_3}\,q_1^2\, \cI_A[1,1,1,0,1]\;, \nonumber \\
 &M_{13}^A = \epsilon^2\sqrt{r_3}\,q_2^2\, \cI_A[1,1,0,1,1]\;, &&M_{14}^A = \epsilon^2q_2^2\,\omega_2\, \cI_A[1,0,1,1,1]\;, \nonumber \\
 &M_{15}^A = \epsilon^2\omega_1\,\omega_2\, \cI_A[0,1,1,1,1]\;, &&M_{16}^A = \epsilon^2\sqrt{\Delta_5}\, \cI_A[1,1,1,1,1][\mu_{11}]\;. \nonumber
\end{align}
Here $\cI_A[1,1,1,1,1][\mu_{11}]$ denotes the insertion of the loop-momentum dependent numerator,
\begin{equation}
 \mu_{11} = \frac{G(\ell,u_1,u_2,q_1,q_2)}{G(u_1,u_2,q_1,q_2)}\;,
\end{equation}
for the pentagon integral that effectively shifts it to $d=6-2\epsilon$
dimensions.  The remaining classical integral families for the topologies shown
in \cref{fig:topologies} can be conveniently obtained through,
\begin{align}
 \cI_B \equiv \left.\cI_A\right|_{u_1 \to -u_1}\;, \qquad
 \cI_C \equiv \left.\cI_A\right|_{u_2 \to -u_2}\;, \qquad
 \cI_D \equiv \left.\cI_A\right|_{(u_1,u_2) \to -(u_1,u_2)}\;.
\end{align}
The same choice of basis (\ref{eqn:pure}) forms a pure basis also for $\cI_B$,
$\cI_C$ and $\cI_D$. These four integral families are sufficient to cover all
occurring integrals in the classical limit. The reason for the reduced set of
topologies is that in the limit $\bm\to\infty$ propagators can become linearly
dependent and can be reduced via partial fractioning. One example for such a
case is the topology shown in \cref{fig:topologyPF}, in which an external
graviton emission from massive internal lines is present.

We now turn to the integrals required in the classical waveform observable,
which combines classical integrals and classical limits of cut-integrals.
Before the classical limit is taken we add the Cutkosky-cut contribution in the KMOC
observable to all integrals (\ref{eqn:Iexp}). Among the integrals topologies
$A$, $B$, $C$ and $D$, only the family $D$ has a non-vanishing cut in the
$(p_1^\prime+p_2^\prime)$-channel, which leads, according to
ref.~\cite{Caron-Huot:2023vxl}, to,
\begin{equation}
\cI_D(\vec{\nu}) + \textrm{Cut}_{1'2'}\,\cI_D(\vec{\nu}) 
	= \cI_D(\vec{\nu}) -2i\textrm{Im}\left[\cI_D(\vec{\nu})\right] 
	= \left[\cI_D(\vec{\nu})\right]^*\;.
\end{equation}
After the classical limit is taken, we work with the complex conjugate
integrals of family $D$.
Given the scaling limit of integral coefficients the integrals appear in
particular linear combinations.  We define such linear combinations of master
integrals and provide analytical results that are valid in the physical region
of phase space given by \cref{eqn:physregion}. We follow
ref.~\cite{Caron-Huot:2023vxl} and define the linear combinations by,
\begin{equation}
 \cI^{\sigma_1,\sigma_2}(\vec{\nu}) \equiv \cI_A(\vec{\nu}) + (-1)^{\nu_2}\sigma_1~\cI_B(\vec{\nu})
 + \sigma_2 (-1)^{\nu_5}~\cI_C(\vec{\nu}) + \sigma_1\sigma_2(-1)^{\nu_2+\nu_5}\Big[\cI_D(\vec{\nu})\Big]^\star\;,
 \label{eqn:linear_comb}
\end{equation}
where $\sigma_i =(-1)^{j_i}$ picks out the mass scaling
$\bar{m}_1^{j_1}\bar{m}_2^{j_2}$.  The pure basis is grouped analogously, as
the algebraic prefactors are identical across all four topologies. Therefore,
we refer to the linear combinations of \cref{eqn:linear_comb} applied to the
master integrals of \cref{eqn:pure} by,
\begin{equation}
M^{\sigma_1,\sigma_2}_n \;,
\end{equation}
where $n$ refers to the particular master integral. We compute the analytical
expressions for the $M_n^{\pm\pm}$ integrals by numerical fitting. We compute
high-precision numerical values for all master integrals $M_n^X$, with
$X=A,B,C,D$, for all four integral families in the physical region using
auxiliary mass flow method ~\cite{Liu:2017jxz,Liu:2021wks,Liu:2022tji}, as
implemented in the \textsc{AMFlow} package~\cite{Liu:2022chg}. We then combine
the numerical values in the linear combinations of \cref{eqn:linear_comb} and
obtain analytical expressions by fitting on weight-$1$ and $2$ functions via
the PLSQ algorithm provided by \PolyLogTools{}~\cite{Duhr:2019tlz}.  Our
results confirm the expressions of ref.~\cite{Caron-Huot:2023vxl}.

For convenience, we now collect the results for the integral functions
$M_n^{\pm\pm}$ in our conventions. All functions $M_n^{--}$ vanish, while   the
functions,
\begin{align}
M_n^{++} \neq 0 \;,
\end{align}
drop out in the classical limit because of vanishing integral coefficients.
The non-vanishing bubble integrals are,
\begin{align}
M_3^{-+} &= \epsilon\Big[-4i\pi\Big] + \epsilon^2\Big[4\pi^2 + 8i\pi\log(-2\omega_1)\Big] + \mathcal{O}(\epsilon^3)\;, \\
M_4^{+-} &= \epsilon\Big[-4i\pi\Big] + \epsilon^2\Big[4\pi^2 + 8i\pi\log(-2\omega_2)\Big] + \mathcal{O}(\epsilon^3)\;.
\end{align}
The non-vanishing triangle integrals are,
\begin{align}
M_5^{-+} &= \epsilon^2\Big[-2\pi^2\Big] + \mathcal{O}(\epsilon^3)\;, \\
M_6^{-+} &= \epsilon^2\left[-2\pi^2 -2i\pi\log\left(\frac{\sqrt{r_4}-\omega_1}{\sqrt{r_4}+\omega_1}\right) \right] + \mathcal{O}(\epsilon^3)\;, \\
M_7^{+-} &= \epsilon^2\left[-2\pi^2 -2i\pi\log\left(\frac{\sqrt{r_5}-\omega_2}{\sqrt{r_5}+\omega_2}\right) \right] + \mathcal{O}(\epsilon^3)\;,  \\
M_8^{-+} &= \epsilon\Big[-i\pi\Big] + \epsilon^2\left[\pi^2  - i\pi\log\left(\frac{y^2-1}{\omega_2^2}\right)\right] + \mathcal{O}(\epsilon^3)\;, \\
M_8^{+-} &= \epsilon\Big[-i\pi\Big] + \epsilon^2\left[\pi^2  +2i\pi\log(y+\sqrt{r_3})- i\pi\log\left(\frac{y^2-1}{\omega_2^2}\right)\right] + \mathcal{O}(\epsilon^3)\;, \\
M_9^{+-} &= \epsilon^2\Big[-2\pi^2\Big] + \mathcal{O}(\epsilon^3)\;, \\
M_{10}^{-+} &=  \epsilon\Big[-i\pi\Big] + \epsilon^2\left[\pi^2  +2i\pi\log(y+\sqrt{r_3})- i\pi\log\left(\frac{y^2-1}{\omega_1^2}\right)\right] + \mathcal{O}(\epsilon^3)\;, \\
M_{10}^{+-} &= \epsilon\Big[-i\pi\Big] + \epsilon^2\left[\pi^2  - i\pi\log\left(\frac{y^2-1}{\omega_1^2}\right)\right] + \mathcal{O}(\epsilon^3)\;. 
\end{align}
The non-vanishing box integrals are,
\begin{align}
 M_{11}^{-+} &= \epsilon\Big[-i\pi\Big]+\epsilon^2\left[\pi^2 - i\pi\log\left(\frac{q_2^2}{4\omega_1^2q_1^2}\right)\right]+ \mathcal{O}(\epsilon^3)\;, \\
 M_{12}^{-+} &= \epsilon\Big[i\pi\Big] + \epsilon^2\left[\pi^2 - i\pi\log\left(\frac{q_1^4(y^2-1)}{\omega_2^2}\right)\right] + \mathcal{O}(\epsilon^3)\;, \\
 M_{12}^{+-} &= \epsilon\Big[i\pi\Big] + \epsilon^2\left[\pi^2 + 2i\pi\log(y+\sqrt{r_3}) - i\pi\log\left(\frac{q_1^4(y^2-1)}{\omega_2^2}\right)\right] + \mathcal{O}(\epsilon^3)\;,\\ 
 M_{13}^{-+} &= \epsilon\Big[i\pi\Big] + \epsilon^2\left[\pi^2 +2i\pi\log(y+\sqrt{r_3})- i\pi\log\left(\frac{q_2^4(y^2-1)}{\omega_1^2}\right)\right] + \mathcal{O}(\epsilon^3)\;, \\
 M_{13}^{+-} &= \epsilon\Big[i\pi\Big] + \epsilon^2\left[\pi^2 - i\pi\log\left(\frac{q_2^4(y^2-1)}{\omega_1^2}\right)\right] + \mathcal{O}(\epsilon^3)\;, \\
 M_{14}^{+-} &=\epsilon\Big[-i\pi\Big]+ \epsilon^2\left[\pi^2 - i\pi\log\left(\frac{q_1^2}{4\omega_2^2q_2^2}\right)\right]+ \mathcal{O}(\epsilon^3)\;,\\
 M_{15}^{-+} &= \epsilon\Big[\frac{i\pi}{2}\Big] + \epsilon^2\left[-\frac{\pi^2}{2} + i\pi\log(y+\sqrt{r_3}) - \frac{i\pi}{2}\log\left(4\omega_2^2\right)\right] + \mathcal{O}(\epsilon^3)\;, \\
 M_{15}^{+-} &= \epsilon\Big[\frac{i\pi}{2}\Big] + \epsilon^2\left[-\frac{\pi^2}{2} + i\pi\log(y+\sqrt{r_3}) - \frac{i\pi}{2}\log\left(4\omega_1^2\right)\right] + \mathcal{O}(\epsilon^3)\;. 
\end{align}
The pentagon integrals contribute only to higher orders in $\epsilon$,
\begin{align}
 M_{16}^{\pm\pm} &= \mathcal{O}(\epsilon^3)\,,
\end{align}
and are not required for the one-loop results.

To obtain the necessary integrals in the physical normalization, a factor of,
\begin{equation}
 c_N = \frac{i}{(4\pi)^2} \Big(4\pi \mu^2 e^{-\gamma_E} \Big)^{\epsilon}\,,
\end{equation}
needs to be included.

The final result for the spectral waveform is written in terms of linearly
independent special functions $f_i$ that we collect here for completeness.
The functions,
\begin{align}\label{eqn:fbasisfin}
 &f_1 = i\pi\;, &&f_2 = \frac{i\pi}{\sqrt{y^2-1}}\;, \\
 &f_3 = \frac{2\log\left(\frac{\sqrt{\omega_1^2-q_2^2}-\omega_1}{\sqrt{-q_2^2}}\right)-i\pi}{\sqrt{\omega_1^2-q_2^2}}\;, && f_4 = \frac{-i\pi}{\sqrt{-q_1^2}}\;, \\
 &f_5 = \log\left(\frac{\omega_2^2}{\omega_1^2}\right)\;, && f_6 = \log\left(4\right)\;, \\
 &f_7 = \log\left(\frac{q_1^2}{q_2^2}\right)\;, && f_8 = \frac{\log\left(y+\sqrt{y^2-1}\right)}{\sqrt{y^2-1}}\;, \\
 &f_9 = \gamma_E - \log(\pi)\;, && f_{10} = 1\;, \\
 &f_{11} = \frac{2\log\left(\frac{\sqrt{\omega_2^2-q_1^2}-\omega_2}{\sqrt{-q_1^2}}\right)-i\pi}{\sqrt{\omega_2^2-q_1^2}}\;, && f_{12} = \frac{-i\pi}{\sqrt{-q_2^2}}\;, \\
  &f_{13} = \frac{1}{\sqrt{y^2-1}}\;, &&f_{14} = \frac{\log\left((y^2-1)\frac{\omega_2}{\omega_1}\right)}{\sqrt{y^2-1}}\;, \\
&f_{15} =\frac{\log\left((y^2-1)\frac{\omega_1}{\omega_2}\right)}{\sqrt{y^2-1}}\;, &&f_{16} = \frac{\log\left(\frac{-q_1^2}{\omega_1\omega_2}\right)}{\sqrt{y^2-1}}\;, \\ 
  &f_{17} = \frac{\log\left(\frac{-q_2^2}{\omega_1\omega_2}\right)}{\sqrt{y^2-1}}\;, &&f_{18} = \log\left(y+\sqrt{y^2-1}\right)\;,
\end{align}
span the basis of $\mathcal{M}^{\rm finite}$, while the remaining regulator dependent functions are necessary to express the $\mathcal{M}^{\rm UV}$, $\mathcal{M}^{\rm IR}$ and $\mathcal{M}^{\rm tail}$ parts of the waveform:
\begin{align}
&f_{19} =\log\left(\frac{\omega_1\omega_2}{\mu_{\rm IR}^2}\right)\;, &&f_{20/21} = \frac{\log\left(\frac{\omega_1\omega_2}{\mu_{\rm IR/UV}^2}\right)}{\sqrt{y^2-1}}\;, \\
&f_{22} =\log\left(\frac{\mu^2}{\mu_{\rm IR}^2}\right)\;, &&f_{23/24} = \frac{\log\left(\frac{\mu^2}{\mu_{\rm IR/UV}^2}\right)}{\sqrt{y^2-1}}\;.
\end{align}
%

%--------------------------------------------------------------------------------
\section{IR Divergence of Waveform Observable}
\label{sec:Weinberg}
%--------------------------------------------------------------------------------
Scattering amplitudes of matter coupled to gravity exhibit universal infrared
singularities \cite{Weinberg:1965nx}. Here we collect the known results,
following ref.~\cite{Brandhuber:2023hhy}.  The infrared divergence of a $n$-point
one-loop amplitude is proportional to tree amplitudes, 
\begin{align}
    M^{1-{\rm loop},\vec s}(p_i)&= \frac{1}{\epsilon} W_S(p_i) \, M^{{\rm tree},\vec s}(p_i) \,,
\end{align}
for $n$ particles with momentum $p_i^2=m_i^2$ and mass $m_i$, which may be
vanishing. Total momentum is conserved $p_1 + p_2 = p_1' + p_2' + k$.  For the
main text we consider the five-point processes
(\ref{eqn:procScalar}) and (\ref{eqn:procVector}).

The soft factor $W_S$ arises from soft graviton exchanges between pairs of
distinct scattering particles. The contribution from particle $i$ and $j\neq i$
is given by,
\begin{align}
\beta_{ij}&=\sqrt{1-\frac{m_i^2m_j^2}{(p_i\cdot p_j)^2}}\,,\\
c_{ij}&=\left\{ 
\begin{array}{ll}
 m_i=0\,\, \mbox{or}\,\, m_j=0\,, &\quad  
 \frac{G}{2\pi}\, 4(p_i\cdot p_j)\,,\\
 \mbox{else}\,, & \quad 
 \frac{G}{2\pi} \frac{ (p_i\cdot p_j) (1+\beta_{ij}^2)}{\beta_{ij}}\,,
\end{array}\right.\\
f_{ij}&=\left\{ 
\begin{array}{ll}
 m_i=0\,\,\mbox{or}\,\, m_j=0\,, &\quad  {\rm log}\left| \frac{2 (p_i\cdot p_j)}{\mu_{\rm IR}^2}\right|  - \pi i  \Theta[(p_i\cdot p_j)] \,, \\
\mbox{else}\,, &\quad   {\rm log}\left[\frac{1+\beta_{ij}}{1-\beta_{ij}}\right]- 2\pi i \Theta[(p_i\cdot p_j)] .
\end{array}
\right.
\end{align}
For emission and absorption by the same external line, the imaginary parts are
absent and one has to rearrange the equations.

For the imaginary contribution one considers all pairs of initial-state and all
pairs of final-state particles,
\begin{align}
 \textrm{Im}\Big[ W_S \Big]=\frac{1}{4}\sum_{\substack{i,j=1\\ i\neq j}}^n  c_{ij}\,f_{ij}  =  - \frac{i \pi}{2} ( 2 c_{12} +  2 c_{1'2'} + c_{1'k} + c_{2'k} )\,.
\end{align}
For the waveform observable we need to add the cut in the
$(p_1'+p_2')$-channel from the 1-loop amplitude. This cut is associated to the
function $f_{1'2'}$, which is the triangle function with the external legs
$p_1'$ and $p_2'$.  Before taking the classical limit the cut can be obtained
as the branch cut of $f_{1'2'}$ in the analytic continuation in $s_{1'2'}$, which
evaluates to $-2 i c_{1'2'} \textrm{Im}(f_{1'2'})$.
Adding this cut then yields the imaginary part of the waveform soft factor,
\begin{align}
\textrm{Im}\Big[ \mathcal{W}_S\Big]  = \textrm{Im}\Big[W_S\Big] + \textrm{Cut}_{1'2'}(W_S) =   - \frac{i \pi}{2} ( 2 c_{12} - 2 c_{1'2'} + c_{1'k} + c_{2'k} )\,,
\end{align}
effectively, flipping the sign of the $c_{1'2'}$ contribution.
Plugging in the kinematic expressions and taking the classical limit we obtain, 
\begin{align}
\mathcal{W}_S = W_S + \textrm{Cut}_{1'2'}(W_S) =  i G (\bar m_1 \omega_1+ \bar m_2 \omega_2) 
    \left[ 1 + \frac{y(2y^2-3)}{2(y^2-1)^{3/2}}\right]\,.
\end{align}
We observe that the hyper-classical terms cancel and that the classical
contribution of the real part of the IR divergence vanishes
\cite{Brandhuber:2023hhy} as well.

%--------------------------------------------------------------------------------
\section{Momentum Parameterization}
\label{sec:Computation}
%--------------------------------------------------------------------------------
As discussed in \cref{sec:reconstruction}, we employ functional reconstruction
techniques to obtain analytical expressions for the classical amplitudes. The
necessary numerical values for this procedure are generated using \Caravel{}.

For convenience and efficiency, we work in finite-field ($\mathbb{F}_p$)
arithmetic, therefore we require a $\mathbb{F}_p$ parameterization of the
external momenta and polarization states.  We obtain the parameterization from
rotations and boosts of a generic momentum configuration and use rational
representations of $\sin{}, \cos{}, \sinh{}$ and $\cosh{}$ functions,
\begin{align}
\sin_r(x)=&\frac{2x}{x^2+1}\;,\qquad \cos_r(x)=\frac{x^2-1}{x^2+1} \;,\\
\sinh_r(x)=&\frac{2 x}{x^2-1}\;,\qquad\cosh_r(x)=\frac{x^2+1}{x^2-1} \;,
\end{align}
which fulfill,
\begin{align}
\sin_r^2(x)+\cos_r^2(x)=1\;,\qquad \cosh_r^2(x)-\sinh_r^2(x)=1\;. 
\end{align}
In order to avoid complex-valued helicity states we work in an alternating
metric signature,
\begin{align}
  \eta =\textrm{diag}\{1,-1,1,-1\}\;.
\end{align}
In this signature one can construct Majorana-Weyl spinors, i.e. real-valued
helicity spinors.  This does not imply any loss of generality, if we extract
rational expressions in momentum invariants $s_{ij}=(p_1+p_j)^2$.  However, we
will need to pay attention to the metric convention in boosts and the
phase-space parameterization.

We can construct a momentum parameterization for $p_1,p_2,p_1^\prime,
p_2^\prime$ and $k$ in the all out-going convention, as this is the natural
parameterization used in \Caravel{}.  It is enough to obtain momenta for
$u_1,u_2,q_1$ and $q_2$ to obtain a full set of external momenta by,
\begin{equation}
\begin{split}
 p_1 &= \bm_1 u_1 + \frac{q_1}{2}\;, \qquad p_1^\prime = - \bm_1 u_1 + \frac{q_1}{2}\;,  \\
 p_2 &= \bm_2 u_2 + \frac{q_2}{2}\;, \qquad p_2^\prime = - \bm_2 u_2 + \frac{q_2}{2}\;,  \qquad k = -(q_1+q_2)\;.
 \label{eqn:mom_outgoing}
\end{split}
\end{equation}
We define the velocities $u_1$ and $u_2$ by,
\begin{equation}
 u_1^\mu = \begin{pmatrix} 1 \\ 0 \\ 0 \\ 0 \end{pmatrix}\;, \quad
 u_2^\mu = \begin{pmatrix} \cosh_r(x) \\ \sinh_r(x) \\ 0 \\ 0 \end{pmatrix}\;,
\end{equation}
while the momentum transfer vectors $q_i$ read,
\begin{equation}
 q_1^\mu = r_1 \begin{pmatrix} 0 \\ \sinh_r(t_1) \\ \cosh_r(t_1) \\ 0 \end{pmatrix}\;, \quad 
 q_2^\mu = r_2 \begin{pmatrix} \sinh_r(t_2)\sinh_r(x) \\ \sinh_r(t_2)\cosh_r(x) \\ \cosh_r(t_2)\cosh_r(t_3) \\ \cosh_r(t_2) \sinh_r(t_3) \end{pmatrix}\;.
\end{equation}
The relations between the physical invariants and these parameterizations are
given by,
\begin{equation}
\begin{split}
 y &= u_1\cdot u_2 = \cosh_r(x)\;, \qquad q_1^2 = r_1^2 \;, \qquad  q_2^2 = r_2^2\;, \\
 \omega_1 &= k\cdot u_1 = - r_2\sinh_r(t_2)\sinh_r(x)\;, \qquad  \omega_2 = k\cdot u_2 = r_1\sinh_r(t_1)\sinh_r(x)\;.
\end{split}
\end{equation}
At this point, the momenta are functions of the above auxiliary
variables,
\begin{equation}
\{r_1,r_2,t_1,t_2,t_3,x\}\;,
\end{equation}
and in the mass parameters $\bm_1$ and $\bm_2$. All of them, except for $t_3$,
are related to the kinematical invariants. The variable $t_3$ is defined
through the on-shellness of the graviton momenta, i.e. $k^2 = 0$, and is given
by
\begin{equation}
 t_3 = \sqrt{\frac{r_1^2 + r_2^2 - 2 r_1r_2\left[\cosh_r(t_1)\cosh_r(t_2) + \cosh_r(x)\sinh_r(t_1)\sinh_r(t_2)\right]}
{r_1^2 + r_2^2 + 2 r_1r_2\left[\cosh_r(t_1)\cosh_r(t_2) - \cosh_r(x)\sinh_r(t_1)\sinh_r(t_2)\right]}}\;.
\end{equation}
With this we have an algebraic parameterization of phase space because of the
square root in the function $t_3$. However, the dependence on the kinematic
invariants is rational. The momenta are then generated as follows,
\begin{enumerate}
 \item Generate random values for $\{r_1,r_2,t_1,t_2,x\} \in \mathbb{F}_p$, such that $t_3$ has a solution in $\mathbb{F}_p$.
 \item Generate random values for $\bm_1$ and $\bm_2$ such that
 \begin{equation}
   \sqrt{\bm_i + \frac{q_i^2}{4}} \in \mathbb{F}_p\;.
 \end{equation}
 \item Construct the momenta according to \cref{eqn:mom_outgoing}.
\end{enumerate}
All square roots are taken in $\mathbb{F}_p$, i.e. for $r=\sqrt{s}$ one
searches a number $r$ in $\mathbb{F}_p$ with,
\begin{equation}
r^2 \textrm{~mod~} p =s\;,
\end{equation}
This number cannot always be found, as the field is not algebraically complete.
However, squares are frequent (approx.~50\% chance) in finite fields and random
sampling of input parameters until all square roots can be taken, is
sufficiently efficient.  

%--------------------------------------------------------------------------------
\bibliographystyle{JHEP}
\bibliography{main.bib}
%--------------------------------------------------------------------------------
\end{document}